\documentclass[floatfix, aps,prx,twocolumn,unsortedaddress,nofootinbib,superscriptaddress,amsfonts,amsmath,amssymb]{revtex4-2}

\usepackage{graphicx}
\usepackage[usenames,dvipsnames, pdftex]{xcolor}

\usepackage{mathrsfs}
\usepackage{mathtools}
\usepackage[mathscr]{eucal}
\usepackage{ulem} 
\usepackage[colorlinks, breaklinks, 
linkcolor=NavyBlue,
citecolor=NavyBlue,
urlcolor=NavyBlue]{hyperref}          

\usepackage{multirow}
\usepackage{tabularx}
\usepackage{booktabs}
\usepackage{comment}
\usepackage{url}

\newcommand{\nn}{\nonumber}

\newcommand{\lab}{\{ }
\newcommand{\rab}{\} }

\newcommand{\eqq}[1]{\begin{align}#1\end{align}}

\newcommand{\msr}{\mathscr}
\newcommand{\zee}{{z_\text{ee}}}

\newcommand{\SB}[1]{{\color{black} #1}}
\newcommand{\ishita}[1]{{\color{black} #1}}
\newcommand{\mS}{{\mathscr S}} 

\newcommand{\xiloc}{\xi_\text{sp}}
\newcommand{\stwo}{\msr{S}_2}

\newcommand{\mI}{{\msr I}}

\newcommand{\mSe}{\msr{S}_\mathrm{e}}
\newcommand{\aI}{\overline{\msr{I}}}
\newcommand{\aSe}{\overline{\msr{S}}_\mathrm{e}} 

\newcommand{\mF}{{\msr{F}}}
\newcommand{\aF}{\overline{\msr{F}}}
\newcommand{\mP}{{\msr{P}}}
\newcommand{\mfq}{{\mathfrak q}}
\newcommand{\mSn}{\mS_\mathrm{N}}
\newcommand{\aSn}{\overline{\mS}_\mathrm{N}}

\begin{document}

\title{The internal clock of many-body delocalization}

\author{Ferdinand Evers}
\affiliation{Institute of Theoretical Physics, University of Regensburg, D-93053 Regensburg, Germany}

\author{Ishita Modak}
\affiliation{Department of Physics, Indian Institute of Technology Bombay, Mumbai 400076, India}

\author{Soumya Bera}
\affiliation{Department of Physics, Indian Institute of Technology Bombay, Mumbai 400076, India}

\date{\today}

\begin{abstract}
After a decade of many claims to the opposite, there now is a growing consensus that generic disordered quantum wires, e.g. the XXZ-Heisenberg chain, do not exhibit many-body localization (MBL) - at least not in a strict sense within a reasonable window of disorder values $W$. Specifically,  computational studies of short wires exhibit an extremely slow but unmistakable flow of physical observables  with increasing time and system size (``creep") that is consistently directed away from (strict) localization. 
Our work sheds fresh light on  delocalization physics: Strong sample-to-sample fluctuations indicate the absence of a generic time scale, i.e. of a naive  ``clock rate"; however,  
the concept of an ``internal clock" survives, at least in an ensemble sense.
Specifically, we investigate the relaxation of the imbalance $\mathscr{I}(t)$ and its temporal fluctuations $\mathscr{F}(t)$, the entanglement and Renyi entropies, $\mathscr{S}_{\mathrm{e}}(t)$ and $ \mathscr{S}_2(t)$, in a 1D system of interacting disordered fermions. 
We observe that adopting  $\mathscr{S}_{\mathrm{e}}(t), \mathscr{S}_2(t)$ as a measure for the internal time per sample reduces the sample-to-sample fluctuations but does not eliminate them. However, a (nearly) perfect collapse of the average $\overline{\mathscr{I}}(t)$ and $\overline{\mathscr{F}}(t)$ for different $W$ is obtained when plotted against $\overline{\mathscr{S}}_{\mathrm{e}}(t)$ or $\overline{\mathscr{S}}_2(t)$, indicating that the average entropy appropriately models the ensemble-averaged  internal clock. 
We take the tendency for  faster-than-logarithmic growth of $\overline{\mathscr{S}}_{\mathrm{e}}(t)$ together with smooth dependency on $W$ of all our observables within the entire simulation window as support for the cross-over scenario, discouraging an MBL transition within the traditional parametric window of computational studies.
\end{abstract}

\maketitle
\section{Introduction}

Many-body localization (MBL) is a spatio-temporal phenomenon that is believed to exist in interacting fermion systems at strong enough disorder~\cite{Basko2006, Gornyi2005, Oganesyan2007, Nandkishore2015, Bera2015, Devakul2015, Prelovsek-review-2017,AbaninBloch-Review-2018,AletReview2018}. 
It manifests itself as a strongly reduced (preMBL) or even fully inhibited  tendency (proper MBL) towards thermalization at large enough disorder strength. The requirement of strong disorder implies that all remnants of relaxation dynamics in the preMBL-regime are necessarily very slow, reminiscent of the familiar behavior of glasses. ``Slowness" originates from the fact that long-time behavior typically is dominated by collective reorganization processes in a highly disordered many-body-energy landscape. The salient collective events are associated with a broad distribution of time scales; there is no single characteristic rate that would lend itself as a measure of time~\cite{YoungRMP86,RiegerProc97}.

We briefly elaborate on the issue of time- and length scales by formulating a scenario: 
Consider an ensemble of disordered quantum wires of a finite length $L$. At strong enough disorder, typical many-body states can be thought about essentially as single (dressed) Slater-determinants constructed from localized single-particle wavefunctions; this is the jest of the celebrated concept of local integrals of motion (LIOM)~\cite{Serbyn2013, Huse2014, Vosk2015, OBrien2016, Imbrie-review2017}. While many, perhaps even most, wires may follow the LIOM-paradigm, a fraction $\mfq(L)$ of the wires will not; they form a thermalizing sub-ensemble that we refer to as ``ergodic bubbles"~\cite{deroeckPRB17, Thiery2017}. A new ensemble of wires with the length $2L$ can be formed by combining samples giving rise to a new fraction $\mfq(2L)$. The evolution of $\mfq(2L)$ with system size has been investigated in RG-studies of toy models~\cite{Vosk2015, Zhang2016, Dumitrescu2018, Dumitrescu2017, Goremykina2018, Morningstar2019, MorningstarPRB20}.

We emphasize two important implications of this rough picture for observations, numerical and experimental, in thermalizing phases when growing the system size at $\mfq(L)\ll 1$: (i) strong finite size effects occur because thermalization sets in only after a certain length-scale have been reached long enough to include an ergodic bubble; (ii) the time scale for relaxation can be exponentially long because it reflects how long it takes a bubble to thermalize an almost localized and large sample region. We refer to this slow relaxation process as `creep'. Since creep emerges from destabilizing an interacting but localized (i.e. LIOM-dominated) sample region via remote thermal bubbles, creep indicates the dominant mechanism for many-body de-localization and is prevalent in the preMBL-regime.

Signals of creep are exponentially long observation times required in numerical and experimental simulations as well as an appreciable dependency of transport-sensitive observables on the simulation volume. 
We interpret the gradual increase of the estimate for the disorder strength beyond which the localization transition was expected to occur as a signature of creep: while early works favored a value $W{\approx} 3.8$~\cite{Luitz2015, MaceMultifractality2018}, later authors reported significant finite size effects and favored larger values $W{\approx 4.5-5.5}$
\cite{Doggen2018, SierantLargeWc20, DoggenRevAnnPhy21}.

\begin{figure}[t]
    \centering
   {\includegraphics[width=1.0\columnwidth]{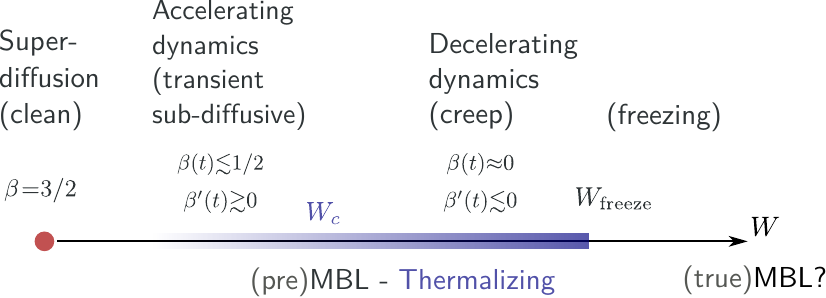}}
   \caption{Schematic phase diagram of the disordered t-V (isotropic Heisenberg) model~\eqref{eq:H} \ishita{ with $V$=1}. Shaded region: thermalizing regime exhibiting (transient) power-laws, e.g., for the decay of imbalance (or time-fluctuations) and growth of entanglement and second Renyi entropy. 
   $W_c{\gtrsim 4.5}$: crossover from accelerating to decelerating dynamics as indicated by the temporal increase/decrease of an effective diffusion exponent $\beta(t)$~\cite{Bera2017,Weiner19}. Beyond $W_\text{freeze}{\approx}10$: fractal dimension of the dynamically active part of the many-body Hilbert space vanishes (freezing)~\cite{NandyPRB21}. This region is largely unexplored in finite size and time numerical studies and hosts the proper MBL phase, if it exists.} 
   \label{f1}
\end{figure}

Creep manifests, in particular, in the spatio-temporal behavior of correlation functions; in the preMBL-regime, they exhibit a spatial decay much slower than exponential with tails slowly increasing with time; 
for a discussion of how creep appears in various observables see \textcite{Weiner19}. 
It is only recently that creep in its diverse manifestations and the resulting implications for the (pre)MBL phenomenology in finite size systems have been appreciated: A weak tendency towards equilibration has been analyzed in the relaxation dynamics even at  large disorder in the local charge density, the sublattice imbalance, and the density matrix~\cite{Bera2017,Weiner19,TaliaPRB19,  
 SierantPRB2021,Morningstar2022,SelsBathPRB22}; the evolution of the entanglement and the number entropy has been investigated ~\cite{SirkerPRL20,SierantPRB2021, SirkerPRB21,MaximilianAP21, Kiefer2022, Luitz2020, Ghosh2022, SirkerComment22, GhoshComment22}; 
attempts have been made to interpret the shift of intersection points (``critical points")
and the flow of the spectral function with increasing system sizes~\cite{Vidmar2020, Panda2019, SierentPRL20, SierantLargeWc20, SuntasPRB20, Polkovnikov2021, AbaninAOP21}.
The status as we see it is summarized in the phase diagram~Fig.~\ref{f1}.

\begin{figure}[t]
    \centering
   \includegraphics[width=1\columnwidth]{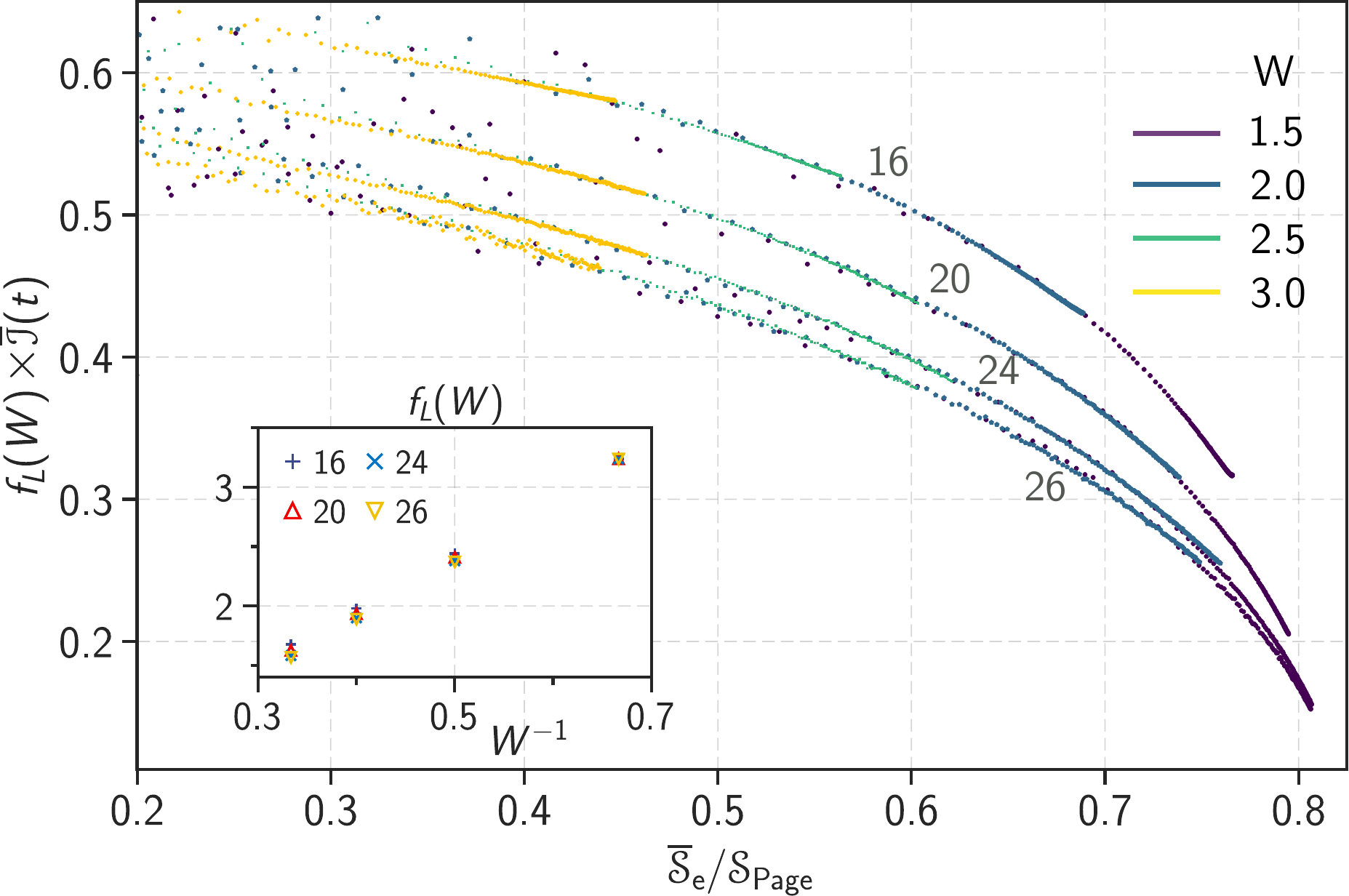}
   \caption{Evolution of the ensemble-averaged imbalance after quenching a product (Ne\'el)-state in a disordered wire of length  $L{=}16,20,24, 26$ at four moderate disorder strengths $W{=}1.5, 2.0, 2.5, 3.0$. Time is measured in terms of the average entanglement entropy in units of the Page-value $\mS_\text{Page}{\coloneqq} \ln 2 (L/2) {-} 1/2$.  
    \ishita{Inset: Scaling factor $f_L(W)$ as a function of $W$. By definition, rescaling the ordinate with $f_L(W)$ collapses imbalance traces for different disorder $W$.
    The factor $f_L(W)$ has been extracted from our data performing a standard scaling analysis. It consists of rescaling the y{-}axis until the best data collapse is observed.
    }
   }
   \label{f2}
\end{figure}

An implication of creep is the absence of characteristic time scales; one rather expects broad distributions of rates characterizing dynamical processes. This poses the question of at which time it is meaningful to compare the dynamical status of two samples that nominally belong to the same ensemble but differ by the specific disorder realization. 
We here pursue the implications of the following hypothesis:  The time evolution in both samples can at least partially be synchronized when introducing an `internal sample time'. We here propose to model the internal time by a form of entropy. While the model has its limitations when adopted on a per-sample level, it works incredibly well for the disorder-averaged `internal ensemble time'. To illustrate the power of the concept, we plot in Fig.~\ref{f2} the average charge imbalance $\aI(t)$ over the entanglement entropy $\aSe(t)$: 
a data collapse is observed in a wide time window for a range of disorder values situated in the (transient) sub-diffusive regime (Fig.~\ref{f1}).

On a more intuitive level, two  considerations motivate us to consider an entropic entity $\mS_L(t)$  as a model of the internal clock of a system with spatial extension $L$: 
(i) The time evolution of the entanglement entropy is, in a sense, the fastest and most stable relaxation process: It is relatively fast, because, unlike the redistribution of, e.g., energy and particle number it is not restricted by a local conservation law: even in the proper MBL phase, one expects a logarithmic-in-time growth of the entanglement entropy, when energy and particle densities have long ceased to relax. The saturation dynamics of physical observables will thus be represented by a plateau when plotted against $\mSe$ in the thermodynamic limit
(ii) On a more formal level, the rate of entropy production can be considered as a generalized driving force for the relaxation of (ensemble-averaged) physical observables~\cite{brenig89}.  
Subjecting the synchronization concept to a further test on a second observable, we apply it to the time fluctuations of the local density, $\aF(t)$. Again an excellent scaling collapse is observed, demonstrating the power of the concept.

Further, we find that $\aF \propto \aSe^{-\rho}$, where the exponent depends on the disorder strength, $\rho(W)$; a similar relationship also holds for typical values of $\mF$. The corresponding exponent functions, $\rho_\text{ave}(W)$ and $\rho_\text{typ}(W)$, reveal opposing trends with increasing disorder strength, in particular, they intersect at a disorder value $W_{\mF}$. Since $W_{\mF}$ is sufficiently close to $W_c$ we take this as fresh evidence in favor of the existence of (at least) two subphases in the thermalizing regime indicated in the phase diagram Fig~\ref{f1}
(`Subphase' is meant here in a weak sense indicating that the thermalizing phase has regions with qualitatively different relaxation behavior that are connected via a cross-over.).

\section{Theoretical setting} 
\label{section2}
\subsection{Model and Method} 
We consider spinless fermions within the $t{-}V$ model 
%
\eqq{
	\msr{H} = & -\frac{1}{2}\sum_{i=1}^{L-1}c_{i}^{\dagger}c_{i+1}+\text{h.c}+\sum_{i=1}^{L} \epsilon_i(n_i-1/2) \nn \\
	& + V \sum_{i=1}^{L-1}(n_i-1/2)(n_{i+1}-1/2), 
	\label{eq:H}
}
where $n_i\coloneqq c_{i}^{\dagger}c_{i}$,  $i$ denotes the site index, $L$ the system size and $V$ the nearest neighbor interaction strength; $V{=}1$ throughout this work.  The on-site potentials, $\epsilon_i$, are uncorrelated and distributed within $[-W, W]$; for comparison, \SB{we present a case of correlated disorder in appendix \ref{appA3}}. 
All calculations are done at half-filling, $N/L{=}1/2$. We study quench dynamics starting from the (product state) charge density wave state $|\Psi\rangle = |1 0 1 0 \ldots\rangle$. \SB{To investigate possible initial-state dependencies, a brief investigation of an alternative initializing state, 
$|\Psi\rangle = |1 1 0 0 \ldots\rangle$, is reported in appendix \ref{appA1}.   }
Time propagation employs the standard Chebyshev expansion of the time evolution operator, which reduces the exponentiation of $\msr{H}$ to several sparse matrix-vector multiplications~\cite{Wei06, Bera2017}. A comparison of the Chebyshev method with exact time evolution is presented in appendix~\ref{app:conv}.

\subsection{Observables} 
In most of this work, we focus on two main observables. For the ``internal clock" we adopt the entanglement entropy
\begin{align}
    \mSe \coloneqq - \text{Tr}_\text{A} \rho_\text{A} \ln \rho_\text{A}, 
\end{align}
where $\rho_\text{A}$ denotes the density operator of subsystem $\text{A}$ as obtained after integrating out the complement of $\text{A}$, i.e. subsystem $\text{B}$. We use an equal partition $L_\text{A}{=}L_\text{B}{=}L/2$. With an eye to the measurement of internal times, we mention that $\mSe(t)$ behaves similarly to the second Renyi-entropy 
$   \mS_2\coloneqq - \ln \text{Tr}_\text{A} \hat \rho_\text{A}^2, 
$
see appendix~\ref{app1}. 
Recently,  $\mSe$~\cite{LukinScience19} as well as the Renyi entropy ${\mS_2}$ have been measured~\cite{BrydgesRenyiScience19} in ion trap experiments.

Our second observable is the sublattice imbalance 
\begin{align}
    \mI(t) \coloneqq \sum_{j}^{L} (-)^j\ n_j(t) 
\end{align}
and derived quantities such as the 
density fluctuations 
\begin{align}
    \mF(t) \coloneqq 1/L \sum_{j=1}^L \lab [n_j(t) - \lab n_j(t) \rab_{\Delta t}]^2 \rab_{\Delta t},
    \label{e4} 
\end{align}
where $\lab\ldots \rab_{\Delta t}$ denotes a sliding time window 
average over a set of given time traces $n_j(t), j=1,\ldots L$, also see ~\cite{NandyPRB21}; 
$\Delta t=12.5$ in our calculations. Observables averaged over 
 different disorder configurations will be denoted by an overline, e.g., $\aSe(t), \aI(t)$. \SB{We also checked our result with a different observable; the detail is provided in App.~\ref{appA2}.}

\section{Clean case - Results \& Discussion} 
Figs. \ref{f3a} and \ref{f4a} display the evolution of $\mSe(t)$ and $\mI(t)$ in the clean system, i.e., $W{=}0$. 

\subsection{Entanglement entropy -  evolution in time and system size} 
\begin{figure}[t]
    \centering
{\includegraphics[width=1\columnwidth]{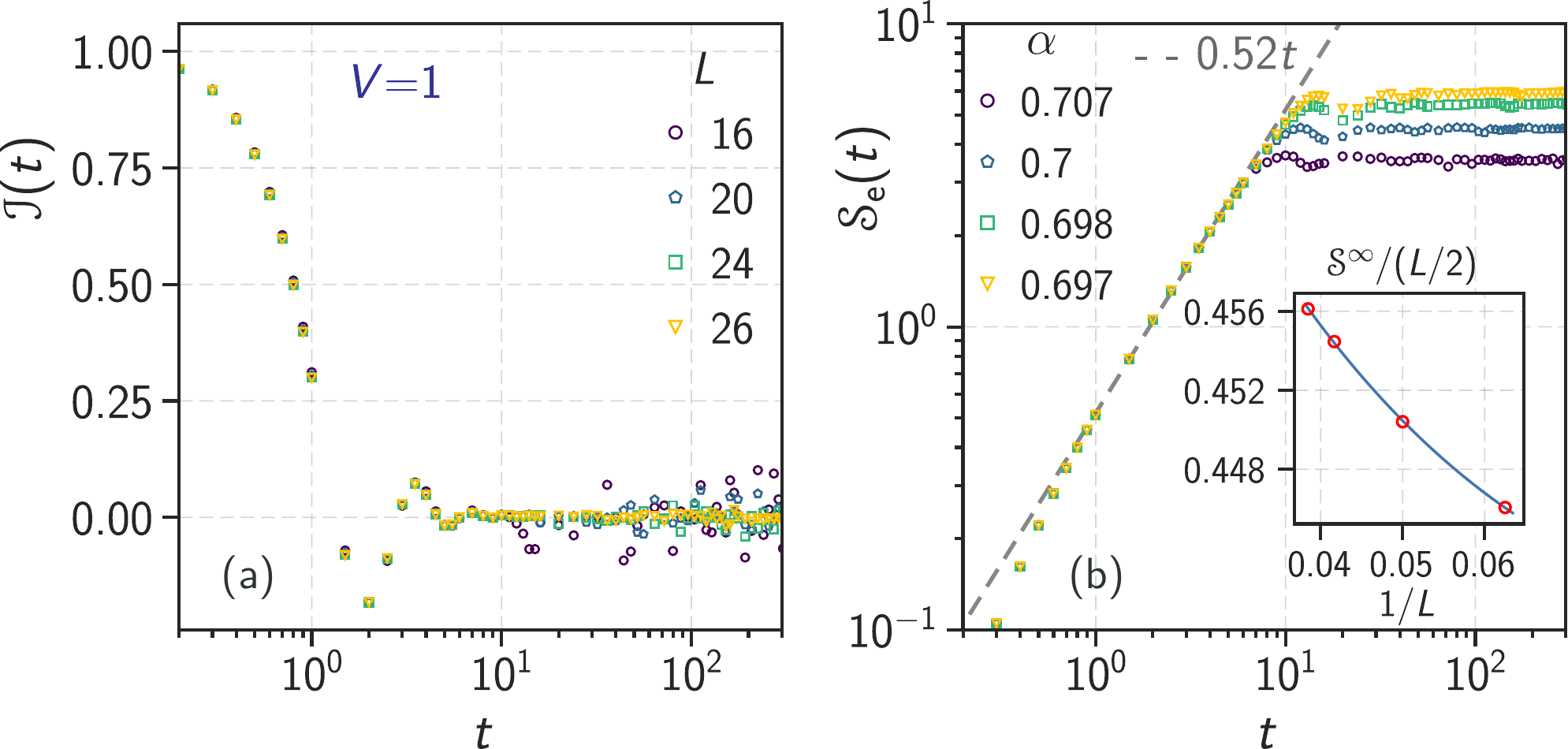}}
   \caption{(a)~Time evolution of the sublattice imbalance, $\mI(t)$  and (b) the entanglement entropy, $\mSe(t)$, after a quench in a clean quantum wire, Eq.~\eqref{eq:H}. $\alpha$ denotes the saturation value of $\mSe$ in units of the Page value $\mS_\text{Page}$ for the respective system sizes. 
   The inset shows the infinite-system size extrapolation of the saturation value, $\mSe^\infty(L)$, as read off the main plot. We take the curvature seen in the data as an indication of corrections to scaling that are non-analytic in $1/L$, possibly logarithmic~\cite{Calabrese_2005}. Fitting the trace to Eq. \eqref{e5} yields
   $a_\text{V}=0.50(1), a_\text{S}=-0.29(4), a=1.8(1).$
   }
   \label{f3a}
\end{figure} 
%

We identify three windows of characteristic times the traces $\mSe(t)$, see Fig. \ref{f3a}b): 
Let the characteristic velocity of pair-excitations be $v$; following the kinetic energy given in Eq. \eqref{eq:H} one expects $v\approx 1$. In the intermediate window, 
$1\lesssim t\lesssim L/2v$,  we have $\mSe(t)=c_\mS t + \text{subleading terms}$,  reflecting the asymptotic behavior expected for the large-system limit, $L\to\infty$, of the ballistic system~\cite{Calabrese_2005}. 
Based on our data Fig. \ref{f3a}(b) we obtain $c_\mS\approx 0.52$. Due to the limited time window and the unknown subleading terms, reliable error bars are difficult to establish. 

Relaxation processes in the short time window occur on rates given by the (inverse) bandwidth; they reflect the local physics of the clean system. As seen in Fig.~\ref{f3a}a), the imbalance decays on these ultra-short time scales. 

At large times, $t\gtrsim L/2v$, the entanglement entropy $\mSe(t)$ reaches a saturation value, $\mSe^\infty(L)\coloneqq \mSe(L;t\to\infty)$, 
that exhibits pronounced system size effects. Its analytical form is well described by 
\begin{align}
\mSe^\infty(L)= a_\text{V} L/2 + a_\text{S}\ln(L/2a) + \ldots
\label{e5} 
\end{align}
where $\ldots$ indicate terms that vanish in the thermodynamic limit and $a$ denotes a microscopic scale that accounts for terms $\mathcal{O}(L^0)$. The first term incorporates the expectation that the entanglement entropy of the ballistic system after saturation is an extensive observable, so the leading behavior is described by a volume law.
Based on quantum field theory, one would expect a prefactor ratio $a_\text{V}/c_\mS=1$, which is consistent with our numerical estimate, $\approx 96$\%, within our extrapolation uncertainties~\cite{Calabrese_2005}.  
An earlier estimate of this ratio, 88\%, has been reported for free fermions (XX model)~\cite{zhao2016}. 
 
The second term in Eq. \eqref{e5} we read as a manifestation of interface (surface) effects: we interpret $L^{d-1}$ in the limit $d{\to}1$ as a logarithm. It accounts for interaction mediated long-range correlations in the one-dimensional bulk phase that lead to anomalously strong surface effects~\cite{Calabrese_2005}. 

For a fully chaotic system, $\mSe^\infty(L)$ is expected to saturate at the Page value, $\mS_\text{Page}\coloneqq \ln(2) L/2 - 1/2$~\footnote{In Ref. \cite{Page1993} it is shown that the average entanglement entropy of random pure states to leading order is of the form 
\begin{align}
    \mS_{m,n} = \ln m - \frac{m^2-1}{2mn} + \ldots \nn
\end{align}
where $n,m$ is the Hilbert space dimensions of the respective subsystems assumed to be large. For the present case, we have $n=m=2^{L/2}$ motivating the definition of $\mS_\text{Page}$ given in the text. 
}. 
Confirming results of earlier authors~\cite{KimPRL13}, we find that clean systems do not exhaust the Page limit, meaning $\alpha <1$, see Fig.~\ref{f3a}b), with  $\alpha(L)\coloneqq \mSe(t\to\infty)/\mS_\text{Page}$ and 
$\alpha(L\to\infty)=a_\text{V}/\ln 2$. For the latter coefficient, we here report an estimate significantly below unity, $\alpha\approx 0.72$.
We interpret this result, $\alpha{<}1$, as a consequence of the fact that the clean model exhibits local conservation laws, such as momentum conservation, that (apparently) can limit the phase-space volume accessible during relaxation as compared to a fully chaotic situation.

\subsection{Clock rate and scaling}
With an eye on the disordered case and its internal clock, we mention that entanglement has been introduced as a suitable time unit before already in the clean case. 
\textcite{Calabrese_2005} observe that the entanglement evolution $\mSe(L;t)$  corresponding to quenches with different initializing states collapse onto a master curve after normalization with respect to the asymptotic value, i.e. when plotting the ratio $\mSe(L;t)/\mSe^\infty(L)$. 
At intermediate times, $1\lesssim t\lesssim L/2v$, $\mSe(t)\propto t$ and therefore the collapse implies $\dot \mSe(t)/\mSe^\infty$ independent of time and the initial condition. In this sense, one might say that the rate of entropy production is controlled by the (saturation) entanglement $\mSe^\infty$.

Taking the idea further, \textcite{KimPRL13} have postulated a scaling form 
\begin{align}
    \mSe(t) \approx \mSe^\infty(L) \ F(t/\mSe^\infty(L))
    \label{e6c}
\end{align}
for a situation in which the system size $L$ rather than the initializing state is varied; 
 $F(x)$ denotes a scaling function. Also here, the saturation value $\mSe^\infty$ sets the rate for entanglement growth. 
 
\begin{figure}[t]
    \centering
{\includegraphics[width=1\columnwidth]{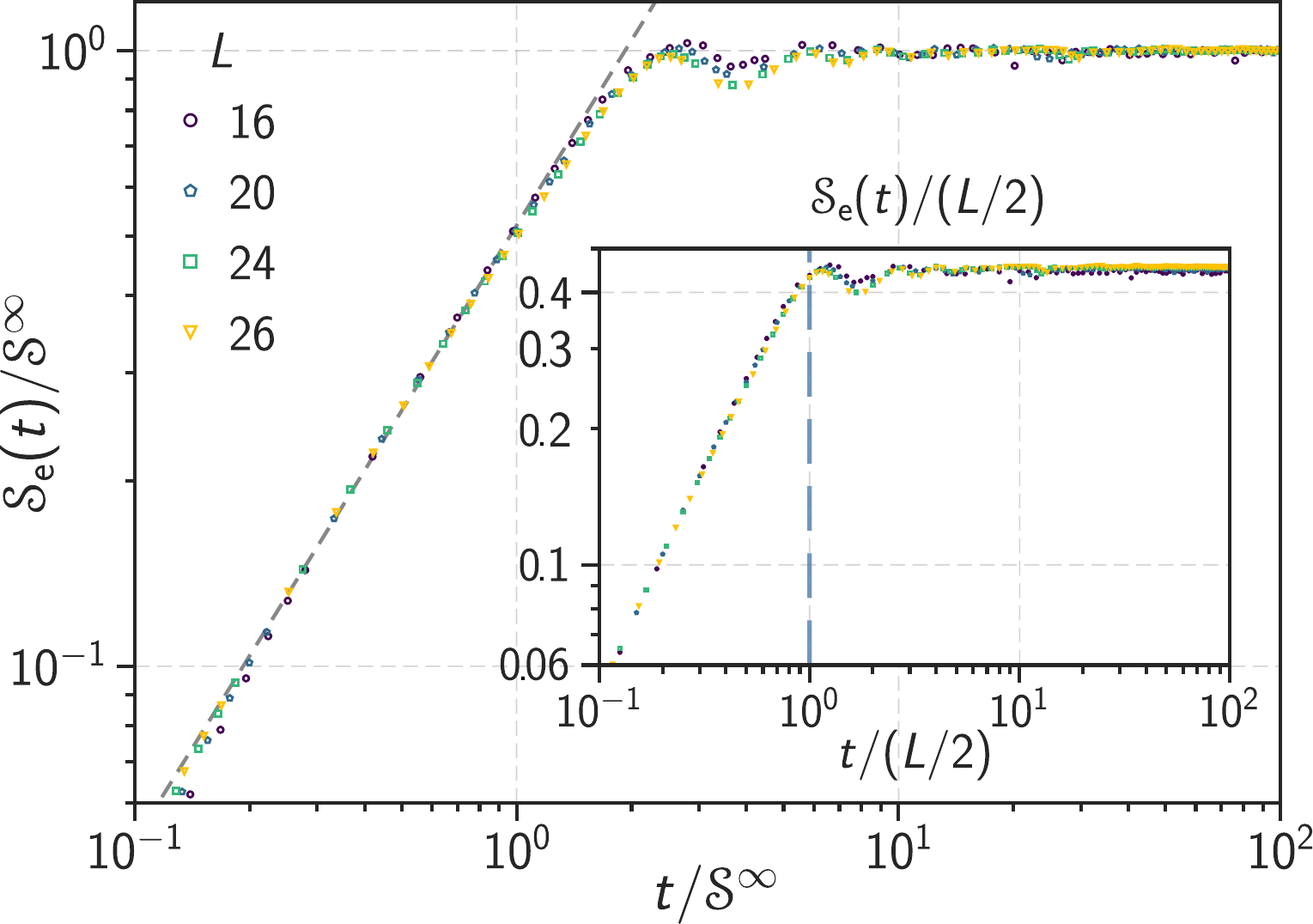}}
   \caption{(Near) scaling collapse of the entanglement entropy after employing the saturation value, $\mSe^\infty(L){\coloneqq}\mSe(L;t\to\infty)$, as a measure of time. Alternatively, the inset employs $L/2v$ for units in order to highlight the finite-size convergence near $t\approx L/2v$ consistent with expectations based on Ref.~\cite{Calabrese_2005}.
   }
   \label{f4a}
\end{figure}

 In Fig.~\ref{f4a} we test the hypothesis \eqref{e6c} and show the scaling function corresponding to the data given in Fig.~\ref{f3a}. While a reasonable collapse is observed, corrections to scaling are appreciable that we tentatively assign to logarithmic corrections, such as displayed in \eqref{e5}. 
Hence, existing reports on quenches in clean systems lend support to the main hypothesis of our work, i.e. that the entanglement entropy can be a good model for an internal clock in 
interacting quantum wires.

\section{Disorder  - Results \& Discussion} 
\subsection{Time evolution of entanglement entropy}
\begin{figure*}[t]
    \centering
   {\includegraphics[width=1\textwidth]{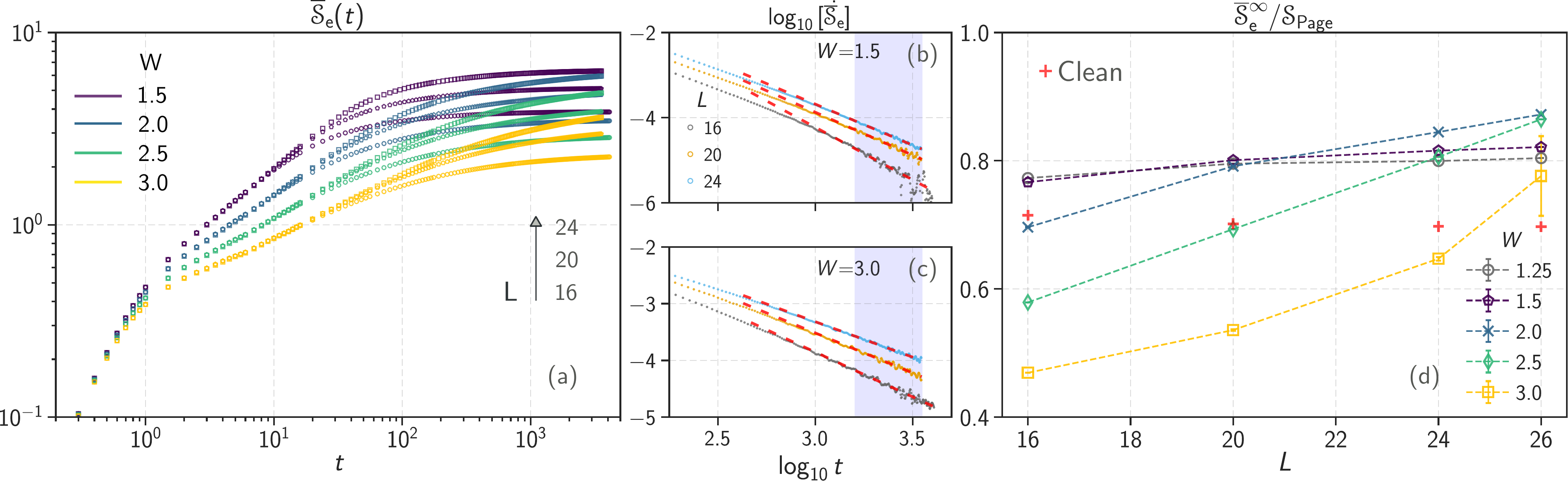}}
   \caption{Ensemble-averaged entanglement entropy $\mSe(L,t)$ for system sizes $L{=}16,20,24$ and four moderate disorder values $W$. (a) Three temporal regimes are identified: ballistic, (sub-)diffusive, and saturating. (b-c): logarithmic time derivative $\dot \mSe$ reveals a (nearly) power-law asymptotics. The shaded region indicates the  window underlying the fits following Eq.~\eqref{e2}. (Parameters in Tab.~\ref{t1}.) (d) Saturation value is seen in the left panel in units of the Page limit $\mS_\text{Page}$ over the linear system size. }
   \label{f4}
\end{figure*} 
We will begin our analysis of disordered systems with the 
ensemble-averaged entanglement entropy $\aSe(t)$ displayed in Figs. \ref{f4}(a) and \ref{f4b}. 
Similar to the clean case, also here the time traces suggest introducing three time regimes:
In the short time regime, all traces (nearly) collapse irrespective of disorder $W$ (ballistic period). This regime we will here leave unattended since disorder effects are (relatively) weak. Our focus is on the sub-diffusive regime (significant disorder effects, system size converged) 
and on the saturation regime (strong effects of disorder and system sizes). 

\subsubsection{Sub-diffusive Regime}
The sub-diffusive regime will eventually display the asymptotic dynamics of bulk systems at large enough $L$ and long enough times, i.e. the thermodynamic limit. Our nomenclature reflects an important result of our investigations also indicated in Fig. \ref{f1}: In the parametric window that we have investigated we consistently observe dynamical signatures, which are expected for the preMBL-regime; MBL proper is never observed. 

Early researchers have characterized the time evolution of observables in the sub-diffusive regime by power-laws,  e.g. $\aSe(t){\propto} t^{1/\zee}$~\cite{Luitz2016, Luitz:2017cp}, 
especially in the range up to $W\simeq 3$ displayed in Fig. \ref{f4}. 
However, as is clearly visible in Fig.~\ref{f4}(a) for the paradigmatic case of entanglement entropy, the putative time window of power-law dynamics has not yet opened up in the range of computationally available system sizes. \footnote{To be extracted from this data is only a lower bound for the dynamical exponent $\zee$, i.e. the slope at the turning point in the (sub-)diffusive regime.}
It thus is apparent that finite size effects in disordered wires are dramatically stronger as compared to the clean case, Fig.~\ref{f3a}. 
\begin{figure}[b]
    \centering
  \includegraphics[width=1\columnwidth]{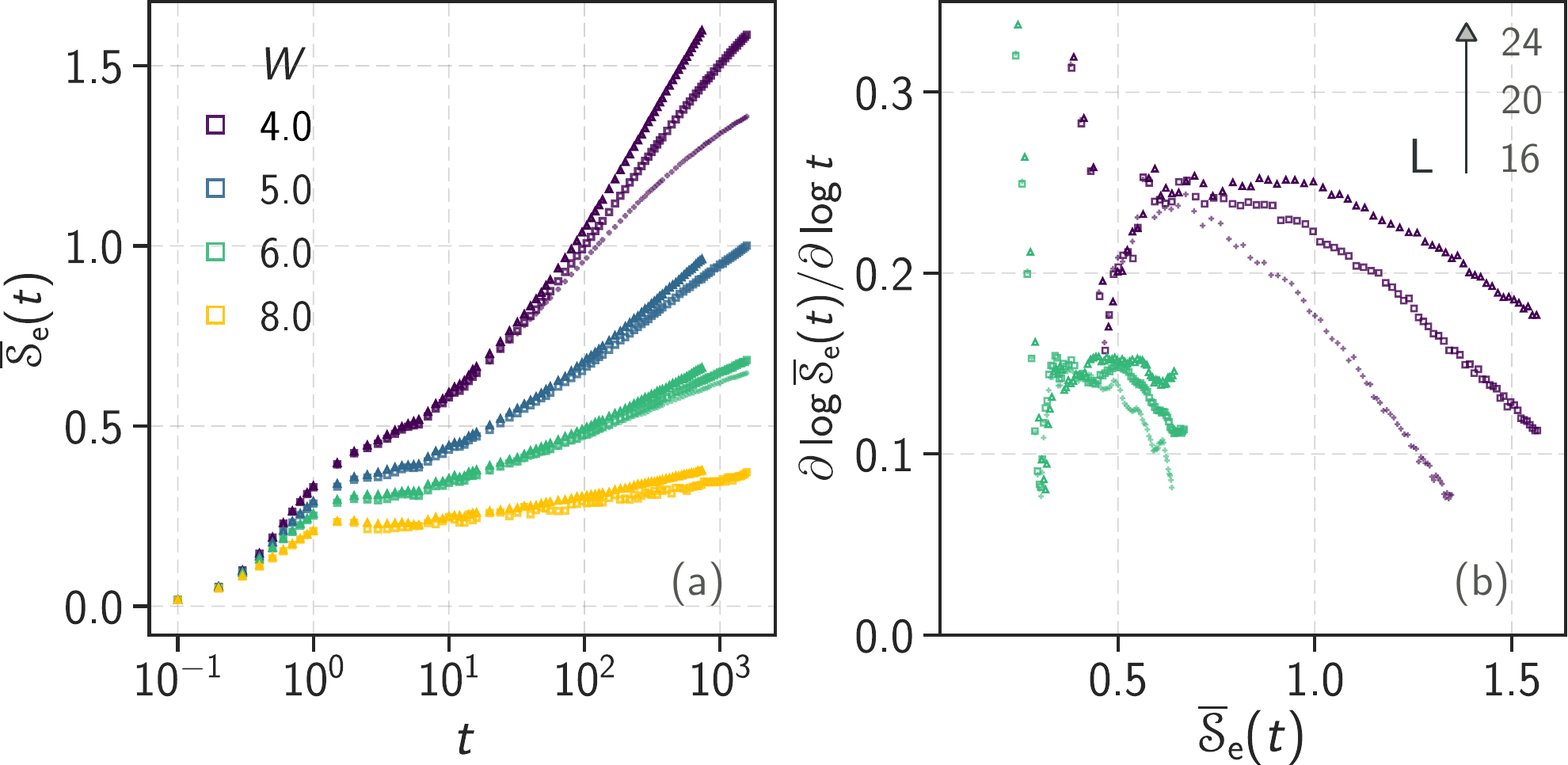}    
\caption{(a)~Traces similar to Fig.~\ref{f4}(a) for disorder in the regime of decelerating dynamics Fig.~\ref{f1}. For $W=4.0,6.0$ system sizes are $L=16,20,24$ and else $L=20,24$. 
(b)~Logarithmic time derivative of $\aSe$ for $W=4.0, 6.0$. The plot highlights the emergence of a power-law increase with effective exponents that quickly grow with the system size $L$. An increase faster than $\ln t$ rules out MBL within the respective disorder regime.  
}
    \label{f4b}
\end{figure}

 The same conclusion also holds at larger disorder in the regime of decelerating dynamics shown in Fig.~\ref{f4b}(a). Notice the
 the left-hand side (lhs) curvature in the intermediate time window where the traces are nearly system-size converged (subdiffusive regime): 
 the crucial observation is that the lhs curvature becomes more pronounced in a larger window of time with ever growing system size. 
 As illustrated in Fig.~\ref{f4b}(b) the curvature signalizes an emergent power law with effective exponents that rapidly increase with the growing system size: a change in system size by 50\% increases the effective exponent at long times by 200\% ($W{=}4$) resp. 50\% ($W{=}6$).
 The phase diagram proposed in Fig.~\ref{f1} is based on the assumption that the trend observed in Fig. \ref{f4b} is indicative of the thermodynamic limit and, therefore, does not reverse at much larger system sizes. While trend reversal towards insulating behavior (``reentrance")  cannot be rigorously excluded in finite-size studies, we don't see any hints pointing toward its existence in numerical work in the MBL context ~
 \footnote{A kind of trend reversal, if only a different one, is observed in the model of regular-random graphs (RRG) \cite{TikhonovRRG16}: this reversal is similar to creep in the sense that the reversed flow in RRG is directed away from the localized to the ergodic regime. Opposed to this, the reentrance described in the main text implies a flow reversal towards localizing behavior. 
 }. 
 In this sense, our result Fig. \ref{f4b} is inconsistent with the expected behavior in the MBL phase, $\aSe(t)\propto \ln t$~\cite{Znidaric2008, Bardarson2012,Serbyn2013a}, and indicates the absence of many-body localization in the respective window of disorder values. 

We mention the implications of our result Fig.~\ref{f4b} for Refs.~\cite{SirkerPRL20,SirkerPRB21}. These authors are quenching random half-filled product states and studying the number entropies $\mSn^{(\alpha)}$
 that they relate  to the second Renyi entropy $\aSn^{(2)} {\sim} \ln \overline{\mS}_2$
and to the entanglement entropy 
$\aSn^{(1)} {\sim} \ln \aSe$ \cite{SirkerPRL20, SirkerPRB21}. Fitting their data in the time domain, they conclude a double-logarithmic growth of the number entropies, $\aSn^{(\alpha)} \sim \ln\ln t$.
These statements imply 
$\aSe(t), \overline{\mS}_2(t) \propto \ln t$, which taken at face value is inconsistent with our claim of growth of $\aSe(t)$ faster than logarithmic, Fig. \ref{f4b}. 
\begin{figure*}[t]
    \centering
   {\includegraphics[width=\textwidth]{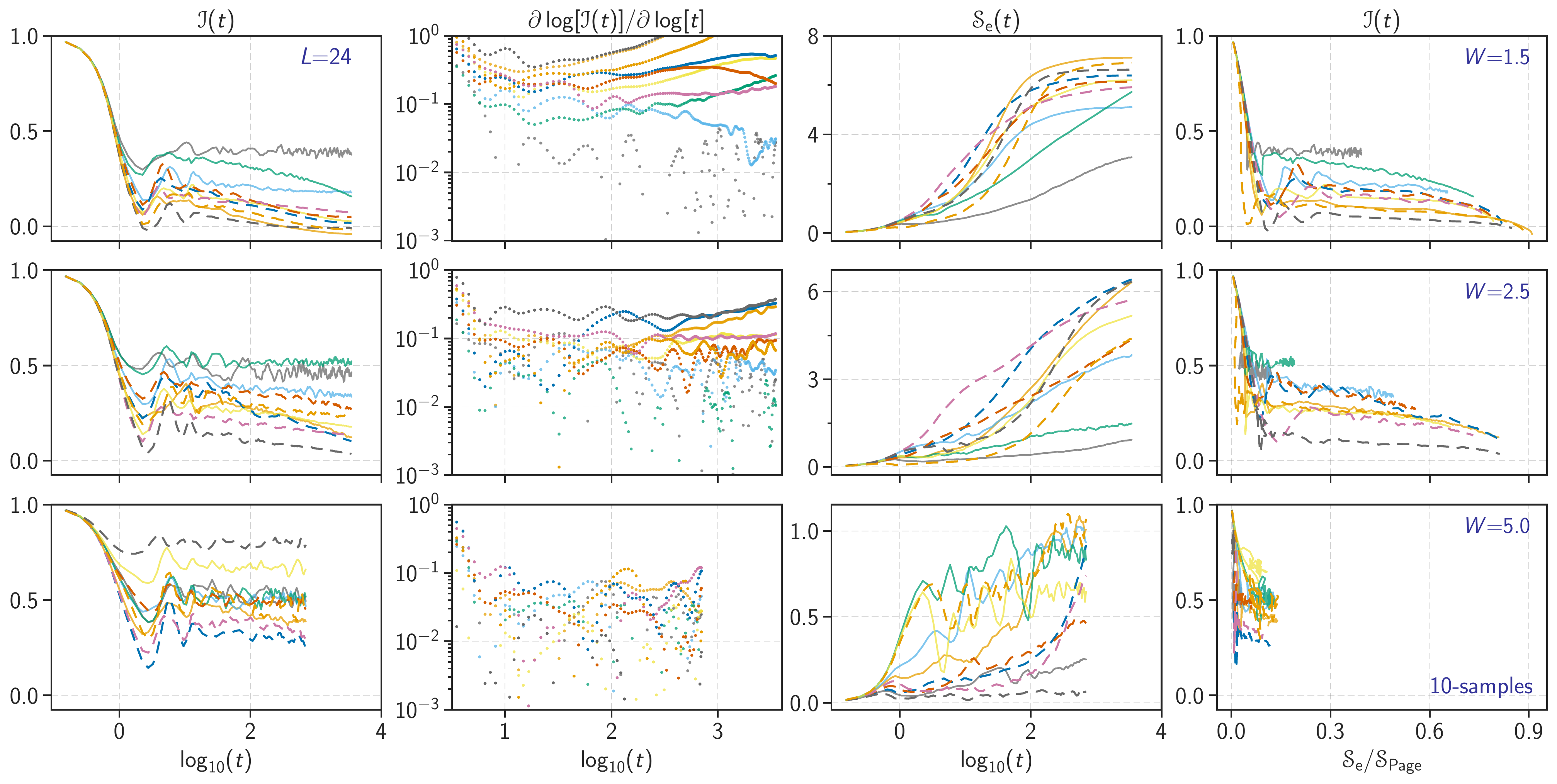}}
   \caption{Plot illustrating extremely strong sample-to-sample fluctuations. Left: Imbalance $\mI(t)$ over simulation time $t$ for $10$ samples and  four disorder values. The plot illustrates the logarithmically wide distribution of $\mI$ at large times.  Second column: Logarithmic derivative (effective exponent $\beta(t)$) of data shown in lhs column. The logarithmically broad distribution of $\beta$ is illustrated. 
   Third column: Corresponding entanglement entropy $\mSe(t)$.
   At large disorder, at a given time $\mSe$ is logarithmically distributed. 
   Right: Imbalance as a function of entanglement $\mS_\mathrm{e}(t)$; sample-to-sample fluctuations are reduced if times with similar values for $\mS_\mathrm{e}(t)$ are compared.
   }
   \label{f5}
\end{figure*} 

We assign this apparent discrepancy to a significant difference in data analysis. 
The $\ln t$ behavior in Refs.~\cite{SirkerPRL20,SirkerPRB21}
has been extracted from time traces obtained from system sizes up to $L{=}24$. In particular, the fit for $L{=}24$ was targeting the largest observation times, similar to our data Fig.~\ref{f4b}(a). Indeed, in this regime also our data is consistent with a logarithmic fitting, as is seen by the lack of curvature in the time traces for $L{=}24$ in Fig.~\ref{f4b}(a) at $t{\gtrsim}10^2$. In this sense, there is no discrepancy here.

However, as is also seen in Fig.~\ref{f4b}(a) the regime of  large times is strongly affected by finite-size effects: with increasing system sizes the traces quickly develop an ever-growing tendency for lhs-curvature that connects smoothly to the lhs-curvature already secured in the sub-diffusive time window. In Fig. \ref{f4b}(b) this flow away from logarithmic towards power-law dynamics is particularly apparent.

The main result of Refs. \cite{SirkerPRL20,SirkerPRB21}, i.e. the absence of MBL-proper in the investigated parameter regime, remains unchallenged here; the reported $\aSe(t)\sim \ln t$ from our point of view represents a lower bound~
\footnote{In our analysis we adopt the perspective that there is no qualitative difference between quenching a Neel-state - as we do - and a random product state as done in Ref.~\cite{SirkerPRL20,SirkerPRB21} for the entanglement dynamics. However, subsequent work noted that averaging over random product states could make it more difficult to extrapolate the long-time dynamics~\cite{MaximilianAP21}.}. 
%

\subsubsection{Saturation regime - disorder enhanced entanglement} 
Confirming earlier results, we readily observe in Fig.~\ref{f4} that also in the disordered situation, $\aSe(t)$ does not exhaust the Page-value, $\mS_\text{Page}$, in the limit of long observation times~\cite{Page93}. To extract the saturation value, $\aSe(L,W)$, we feed the asymptotic form  
\ishita{
\begin{align}
    \aSe(L,W;t) = \aSe^{\infty}(L,W) - c(L,W)/t^{\gamma_L(W)} 
    \label{e2}
\end{align}
}
into a three-parameter fit of the data Fig.~\ref{f4} (a-c). To estimate the amplitude $c(L,W)$ and the exponent $\gamma_L(W)$, we employ the asymptotics of the time derivative, $\dot \aSe_W(L,t)$, reproduced in Fig.~\ref{f4}(b,c). With this input, the saturation value $\aSe^\infty(L,W)$ is fitted adopting \eqref{e2} to the data in the main panel, Fig.~\ref{f4}(a); the fitting parameters are reproduced in the appendix, Tab.~\ref{t1}.

We emphasize that the exponent $\gamma_L(W)$ is an effective one. Close scrutiny of Fig. \ref{f4}(b,c)  reveals a slow time evolution of the slope (exponent) that is nearly but (perhaps) not fully converged within our window of observation times. As a consequence, the fitting parameters contain a systematic error that is not accounted for in the error estimates given in the appendix, Tab.~\ref{t1}. 

The saturation values of the entanglement entropy, $\aSe(L,W)$ are reproduced in Fig. \ref{f4}(d). A striking aspect revealed here is {\it disorder enhanced entanglement}; there is a wide regime in the $L,W$-plane, in which the subsystem-entanglement {\it grows} with the disorder. This is always the case in the limit of weak disorder, 
where it perhaps is less surprising: weak disorder destroys integrability and therefore brings the system closer towards chaotic dynamics, so $\aSe(L,W)$ moves upwards towards the Page limit. This increase has been seen previously in Ref.~\cite{Singh2016}.
Remarkably, the trend appears to continue even to moderate and strong disorder, $1.5\lesssim W \lesssim 3.0$, provided the system size is large enough - as seen from the traces with family parameter $W$ intersecting each other in Fig.~\ref{f4}(d). 

\subsection{Entanglement entropy as an internal clock} 

In this subsection, we explore the potential of entanglement entropy as a model for an internal clock. 
We begin by pointing out a possible advantage of using internal clocks. 
Time traces that display the relaxation of two different transport-related observables both measured in a given sample tend to be correlated. For example consider $\mSe(t)$ and $\mI(t)$: When employing $\mSe(t)$ as a measure of time on a per-sample basis, sample-to-sample fluctuations should reduce considerably, because the slow growth of the first will, in general, indicate strong localization and therefore hints at slow growth of the latter; ergo,  $\mI(\mSe)$ fluctuates less than  $\mI(t)$ between different samples and therefore should be easier to interpret.  The reduction of sample-to-sample fluctuations is indeed seen in  Fig.~\ref{f5}, right most column.  

We elaborate on the qualitative argument. It can be rephrased employing concepts underlying the traditional mode-coupling theory\SB{~\cite{brenig89}}. It  foresees that two slow observables in their time evolution can each follow their own internal clock rate, but only if they are `orthogonal', i.e., sufficiently decoupled. The generic expectation is that orthogonality is an exception; at least  quantities deriving from the same (conserved) field, e.g. the particle density, follow the same clock rate. 
 
Adopting $\mSe$ as a model for the internal clock illustrates a benefit of the concept, i.e. reducing sample-to-sample fluctuations in Fig.~\ref{f5} significantly. At the heart of the concept is the hypothesis that improved models (as compared to $\mSe$) exist, at least in principle, that further reduce sample-to-sample fluctuations, nearly eliminating them in generic observables at large enough $L$. 
 The optimal model for the internal time on a per-sample level still has to be found. However as we will see in the next paragraphs, $\aSe$ is close to an optimal model for the disorder-averaged internal clock, i.e. the `internal ensemble time'.

\subsubsection{Excursion: the arrow of time - progress and regression} 

A conservative interpretation of $\mSe(t)$ as the internal system clock requires that the time evolution of the entanglement entropy exhibits a monotonous growth not only on average but also on a sample-to-sample basis.
This requirement is not obviously met,  because the entanglement entropy is not bound to strict growth in time far enough in non-equilibrium. As seen in Fig.~\ref{f5}, first and third column, 
similar to the charge imbalance~\cite{NandyPRB21} $\mI(t)$, also the time evolution of the entanglement entropy $\mSe(t)$ exhibits strong sample-to-sample fluctuations. 
And indeed, as seen in Fig. \ref{f5}, the third column at strong disorder $W\gtrsim 4$, $\mSe(t)$ becomes highly non-monotonous. Hence, a conservative interpretation of $\mSe(t)$ as an internal system clock will confine itself to moderate disorder $W\lesssim 4$. 

We would like to argue, however, that strict monotony is not a necessary condition for an observable to be a useful measure of time. To this end we recall that the arrow of time is bound to point in a single direction only in the macroscopic limit; relaxing finite-size systems may always exhibit transient periods of entropy shrinkage (`regression'), so that the condition of strict monotony can be released, at least in principle.

\subsubsection{Collapsing averaged imbalance traces -  $\aI(t)$ over $\aSe(t)$} 
\begin{figure}[t]
    \centering
   {\includegraphics[width=0.975\columnwidth]{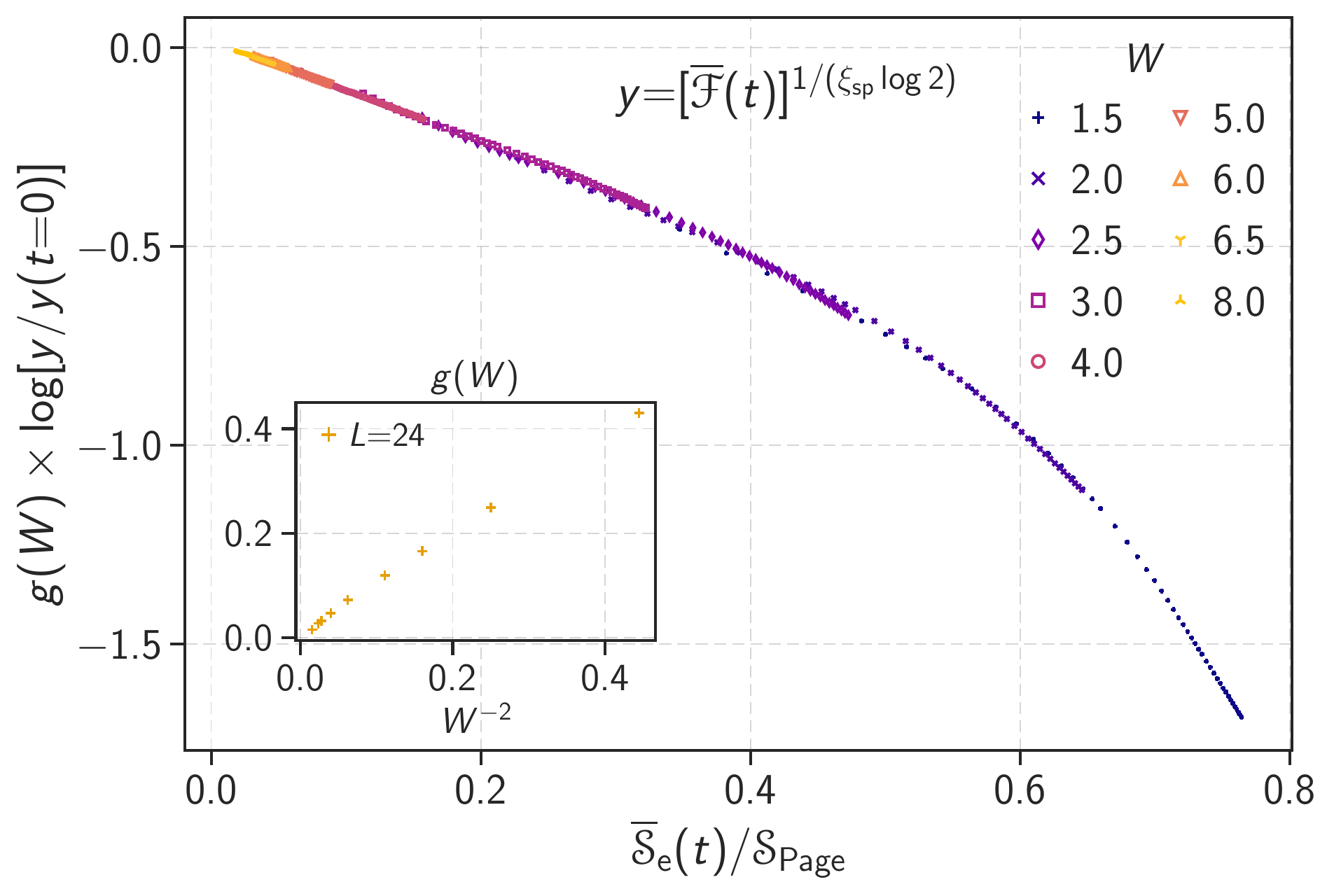}}
   \caption{Approximate collapse to a single master curve of the traces for $\aF(t)$ for different disorder values $W{=}1.5, 2.0, 2.5, 3.0, 4.0, 5.0, 6.0$ displayed in Fig.~\ref{f6a} for $L=24$. \SB{$\xi_\mathrm{sp}$ denotes the non-interacting localization length (further details in Ref.~\cite{Bera2017})}. \ishita{The $\aSe$ traces are taken from Fig.~\ref{f4} and Fig.~\ref{f4b}. Inset: Prefactor $g(W)$ defined by the scaling collapse as a function of $1/W^2$. } }
    \label{fig:dephasing}
\end{figure}

In the spirit of the preceding paragraphs, we adopt $\aSe(t)$ as a legitimate model for the internal clock of the ensemble dynamics also at larger disorder values, even if individual samples display periods of regression. 
In fact, we will continue to bravely use $\aSe(t)$ also at large disorder as displayed, e.g., in Fig. \ref{fig:dephasing}. 

\begin{figure}[t]
   \centering
   \includegraphics[width=0.975\columnwidth]{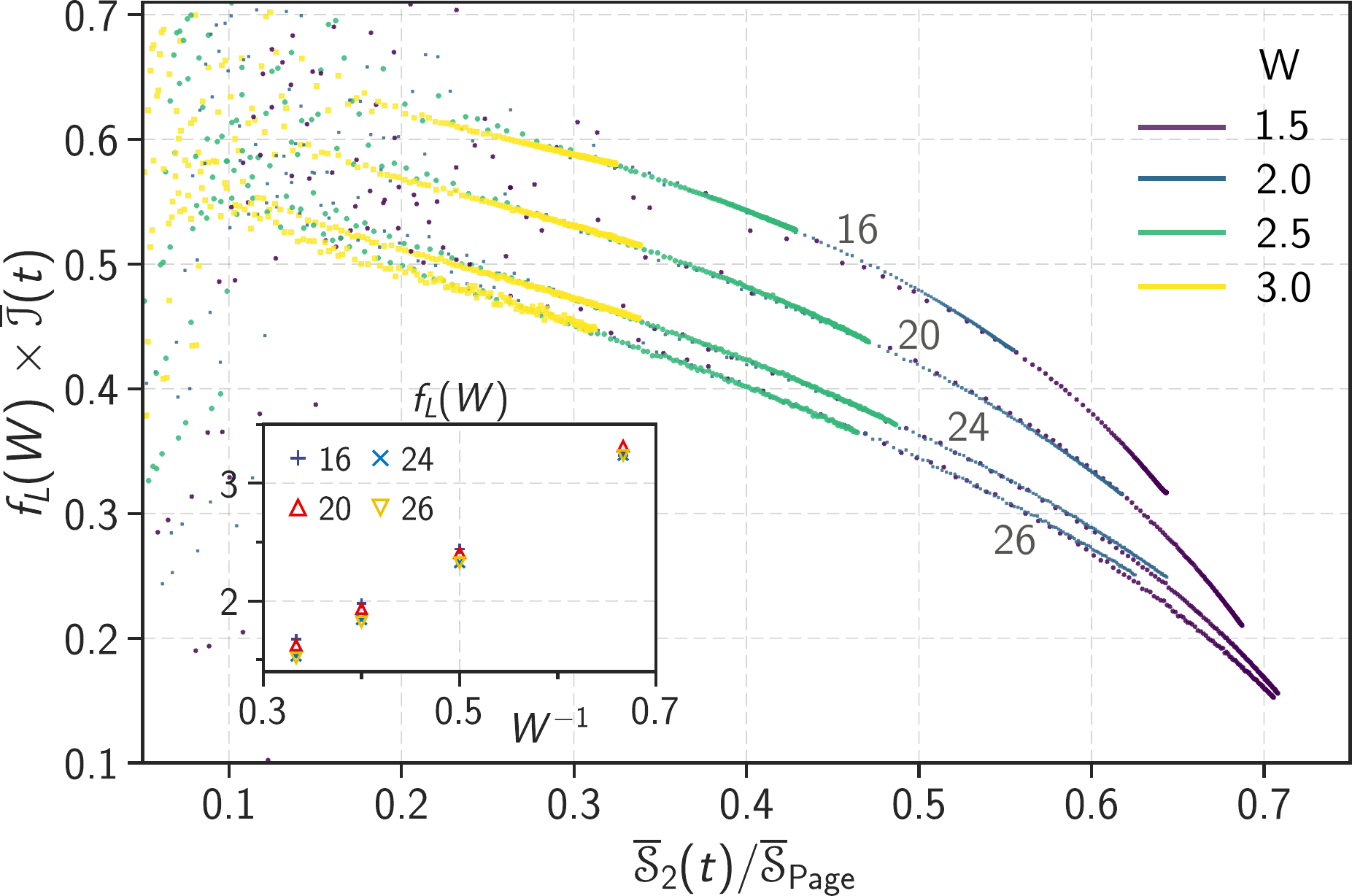}
   \caption{Data from Fig.~\ref{f2} plotted over the second Renyi-entropy. A good collapse is achieved here, also, confirming that $\aSe(t)$ and $\overline{\mS}_2(t)$ model the internal clock with comparable quality. A direct comparison of $\aSe(t)$ and 
   $\overline{\mS}_2$ is given in the appendix, Fig.~\ref{f12}. Parameters:  $L{=}16,20,24, 26$ at four moderate disorder strengths $W{=}1.5, 2.0, 2.5, 3.0$. \ishita{Inset: prefactor $f_L(W)$ as a function of $W$. The factor $f_L(W)$ has been extracted from our data in a similar process as Fig. \ref{f2}.
}}
    \label{f7a}
\end{figure} 

As we have argued, internal clocks are useful to the extent that they reveal similarities in the time evolution of two different samples that are not apparent when using the lab-time $t$. A particularly strong case in favor of internal clocks can be made if similarities can be revealed in the dynamics of systems - or ensembles - that nominally exhibit strong  differences, e.g., in the disorder strength, so that at least `naively' similarities are not expected. 
Such a strong validation of the  ``internal clock" concept has been given with Figs.~\ref{f2} and \ref{f7a}: imbalance traces $\aI(t)$ taken from samples of fixed length $L$, and for disorder varying from the moderate ($W{\approx} 1.5$) to the beginning of the strong disorder regime ($W{\approx} 3$) collapse to a master-curve - within the available window of observations times.

For collapsing $\aI$, rescaling of the abscissa is not required, which reflects the absence of relevant microscopic time scales - at least within the observation window and outside the short-time regime. 
The fact that the data collapse requires a rescaling of the ordinate is not unexpected, but still merits a remark: Our computations operate in the limit of infinite temperature, $T^{-1}{=}0$; the corresponding equilibrium density is spatially homogeneous, so that the equilibrium imbalance vanishes, $\mI{=}0$. Therefore, this observable or quantities derived thereof do not lend themselves to compare to the relaxation behaviour seen in $\aI(t)$. Due to the apparent lack of an obvious scale derived from an equilibrium quantity, it is not inconceivable that the scaling factor of the ordinate, $f_L(W)$ 
(inset of Fig.~\ref{f2}), 
reflects an initial condition, here the N\'eel state, and changes upon other choices. \SB{ The factor $f_L(W)$ has been extracted from our data performing a standard scaling analysis. It consists
of rescaling the y-axis until the best data-collapse is observed. }\\
We mention that at late times when entanglement approaches its saturation value, the master curve exhibits a large - possibly diverging - slope. The implication is that entanglement propagates fast and may saturate long before other physical observables do, which are subject to a local conservation law. 

\subsubsection{Collapsing of density fluctuations --  $\aF$ over $\aSe$} 
 Traditional mode-coupling theory suggests that if $\aSe$ is suitable as an internal clock for $\aI$ then it should also be for related observables that derive from the evolution of density modes. In this spirit, we show in Fig. \ref{fig:dephasing} that also the time evolution of imbalance fluctuations \eqref{e4} can be collapsed employing the entanglement evolution, $\aSe$, as an internal clock.
\SB{In appendix \ref{appA2} we offer preliminary evidence suggesting that a collapse is also achieved for a third variable, i.e. the density auto-correlation function}. 

Since we have already demonstrated the concept of a system-internal clock for $\aI(t)$ in the regime of accelerated dynamics, $W\lesssim 3$, the excellent scaling observed for $\aF$ in this regime is comforting but not, perhaps, surprising. Interestingly, a reasonably good data collapse is seen for $\aF(t)$ even in the regime of large disorder, $W\approx 8$, which we take as an additional encouragement for the concept of internal clocks, here proposed. 

Incidentally, it is implied here that at least with respect to the observable density fluctuations (as seen within our observation time) there is no indication of a phase transition all the way from moderate to strong disorder.

\subsection{Intermediate power-laws in average \& typical $\aF$} 
\begin{figure}[t]
    \centering
   {\includegraphics[width=0.95\columnwidth]{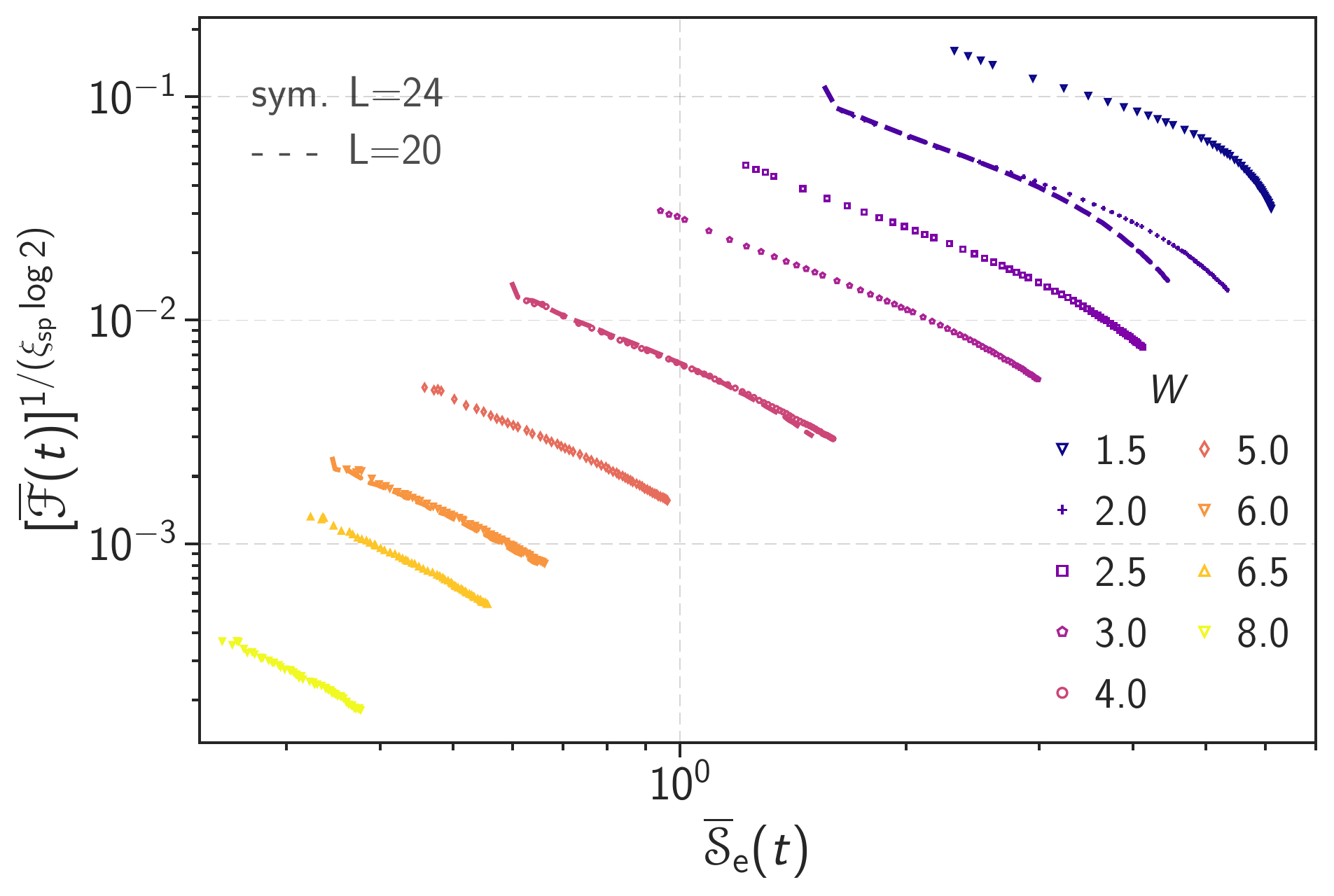}}
   \caption{Ensemble-averaged temporal fluctuations of the imbalance $\mI(t)$ as defined in \eqref{e4}. Following our earlier work [\onlinecite{NandyPRB21}], we have dressed $\aF$ with a power $1/\xi_\text{sp}\log 2$. For selected traces, two system sizes, $L=20~(\mathrm{lines}),24~(\mathrm{symbols})$, are given to expose finite-size effects. 
   }
   \label{f6a}
\end{figure}

The raw data underlying the master curve is displayed in Fig.~\ref{f6a}. The data confirms a trend already observed in our previous work \textcite{NandyPRB21}: the imbalance fluctuations exhibit a rather benign convergence with the system size. We, therefore, are confident to identify for each trace an intermediate regime between short times (plateau region) and longest times (pronounced system-size dependency) that defines an (effective) power law
\begin{align}
\log \aF/\aF_1 &= - (\rho_\text{ave}\ \xiloc \log 2)\log  \aSe
\label{e6a} 
\end{align}
with $\aF_1$ denoting the prefactor; $\xiloc$ is the non-interacting localization length extracted in the long time limit of the second moment of the density-density correlation function at infinite temperature~\cite{Bera2017}.
%
%
The exponent function  $\rho_\text{ave}(W)$ is extracted fitting the data Fig.~\ref{f6a} to the form \eqref{e6a}; an analogous fitting procedure can also be performed for the typical traces of $\mF$ and $\mSe$ yielding $\rho_\text{typ}(W)$.

\begin{figure}[t]
    \centering
   {\includegraphics[width=0.95\columnwidth]{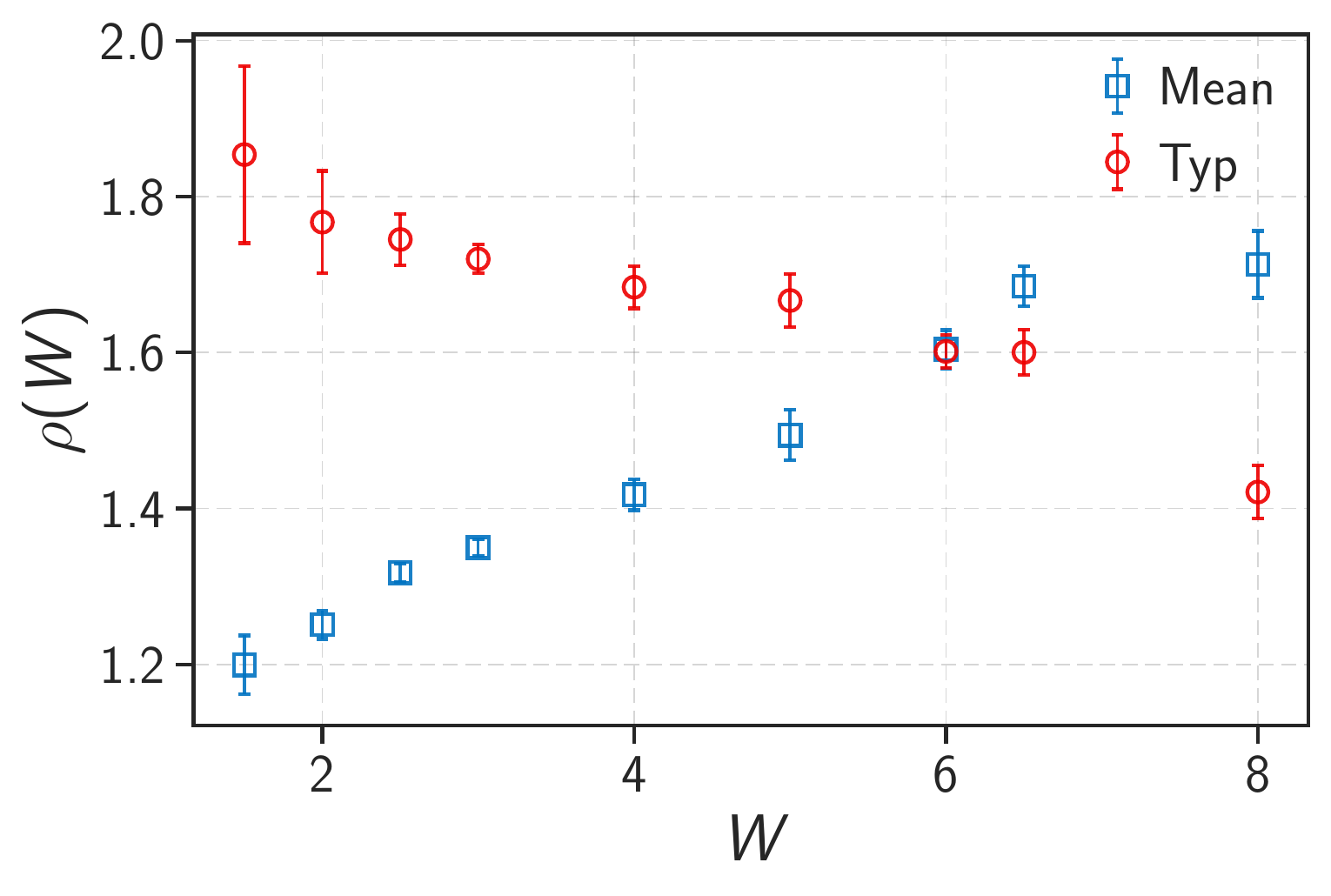}}
   \caption{(Effective) exponents characterizing the evolution of the imbalance fluctuations with the entanglement entropy, 
   $\mF{\propto} \mSe^\rho$, for typical values, $\rho_\text{typ}$, and average values $\rho_\text{ave}$. The intersection point indicates the crossover from accelerated to decelerated dynamics, see Fig.~\ref{f1}.    
   }
   \label{f8}
\end{figure}

The results for both exponents are displayed in Fig.~\ref{f8}. 
We first note that the evolution of the traces with the disorder is rather smooth and slow; there is no indication of a nearby MBL-transition.
\footnote{
One might ask why we expect $\rho(W)$ to exhibit a signature of the transition into the proper MBL phase: the definition of the exponent, 
$\aF\propto \aSe^\rho$, at least partially reflects the fact that $\aF(t)$ and $\aSe(t)$ exhibit a power-law growth at long times (in a large-enough system). At least for the entanglement entropy, $\aSe(t)\propto t^{1/z_\text{ee}}$,  the growth becomes slower than any power, $1/z_\text{ee}\to 0$, so that generically the proportionality  $\aF\propto \aSe^\rho$ should cease to hold. 
} 

Remarkably, the exponent functions display opposing trends with increasing disorder. The typical value follows the {\it naive} expectation,  which is that with the increasing disorder, the interaction-induced damping of temporal fluctuations is less effective, so $\rho_\text{typ}$ has a tendency to decrease. 
While the typical exponent $\rho_\text{typ}(W)$ decreases, the average exponent $\rho_\text{ave}(W)$ increases with disorder and there is, indeed, a common intersection point 
 $W_{\mF}\approx 6$. At weaker disorder, $W{<}W_{\mF}$, rare samples exist that exhibit rather long decoherence times, so in this regime $\rho_\text{typ}{>}\rho_\text{ave}$. In the other regime, $W{>}W_{\aF}$ the situation is reversed and the average is dominated by rare samples in which damping is unusually effective.  Notice that the crossover value is situated near the regime in which the system dynamics changes from accelerated to decelerated, $W_{\aF}\approx W_c$, cf. Fig. \ref{f1}.

 Summarizing, we interpret Fig.~\ref{f8} in conjunction with Fig.~\ref{f4b} as supplying fresh  support in favor of the phase-diagram Fig.~\ref{f1} and its main claim: In the range $W_c,W_{\aF}\approx 3-6$ there is a crossover between two markedly different thermalizing regimes; 
 there is no evidence that the $t-V$-model exhibits a phase transition at disorder values below $W\lesssim 10$.
 We thus advocate the point of view that until recently
  this crossover has been widely misinterpreted in numerical work as indicating an ergodicity-breaking (i.e. MBL-) transition, e.g. in Refs.~\cite{Pal2010, Berkelbach2010, BarLev14, Devakul2015, Luitz2015, Khemani2017, Loic2018, MaceMultifractality2018, SierantLargeWc20, Laflorencie2020, ChandaPRB20, DoggenRevAnnPhy21}.
 Recent numerical work claims that the MBL transition if it exists, occurs at a disorder strength $W^*\gtrsim 20$~\cite{Morningstar2022, SelsBathPRB22}. 
We notice that already at disorder value $W{\approx}3$ the non-interacting localization length $\xi_{\mathrm{sp}}$ is of the order the lattice spacing $a$, 
which implies $\xi_{\mathrm{sp}}(W^*)\ll a$.  
 In this sense, if the MBL transition occurs at all, it for sure takes place at an extremely strong disorder. 
This immediately prompts the question of its physical significance.
 %

\subsection{Sample-to-sample fluctuations and thermalization}
\subsubsection{Distribution of exponents} 
\begin{figure}[t]
    \centering
   {\includegraphics[width=1\columnwidth]{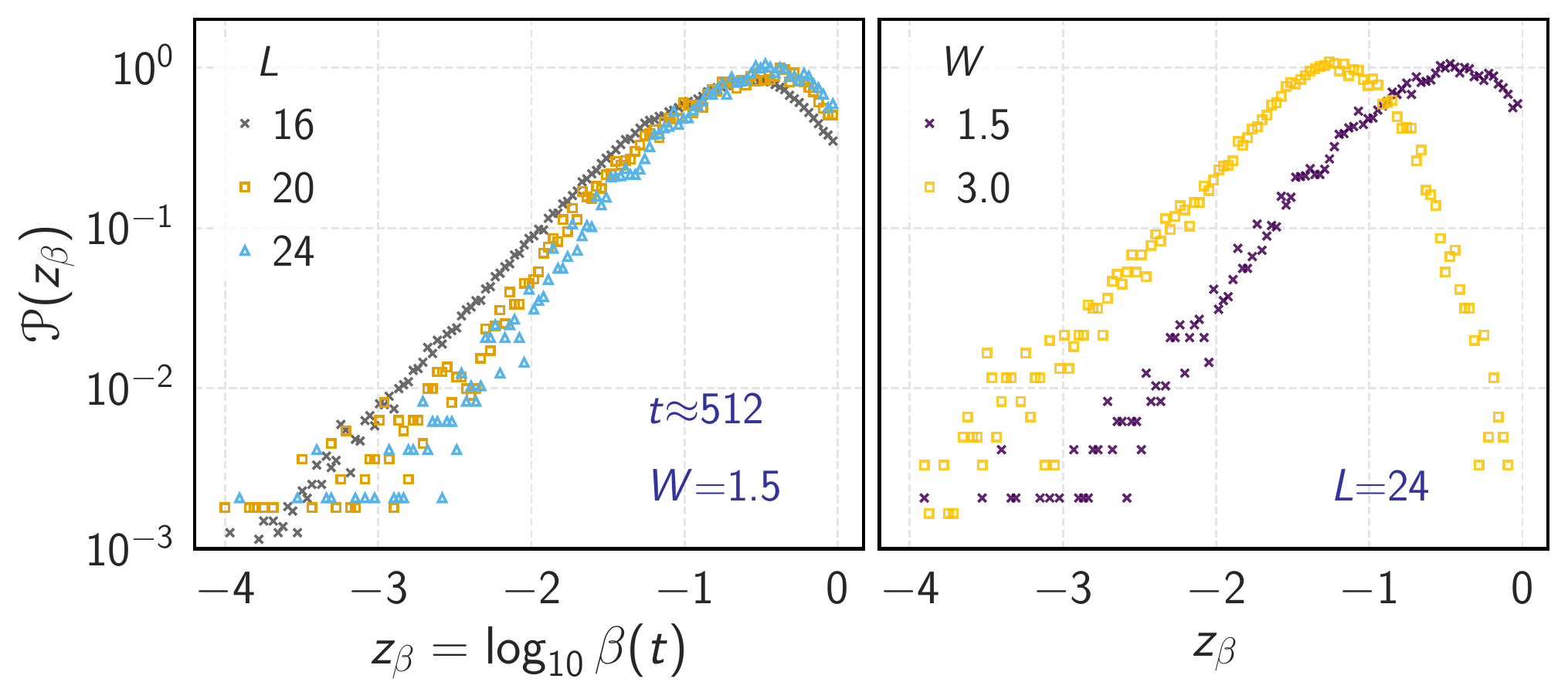}}
   \caption{Distribution function $\mP$ of the effective exponents $\beta(t)$ (Eq. 
   \eqref{e6}) that characterize the long-term behavior of the imbalance $\mI(t)$ after a quench in systems of sizes $L=16,20,24$ and for two disorder values in the (sub-)diffusive regime. }
   \label{f6}
\end{figure}

The importance of sample-to-sample fluctuations is demonstrated in Fig.~\ref{f5}, lhs column. The plot shows the distribution of $\mI$ being logarithmically broad at large observation times already at moderate disorder, $W{=}1.5$. In fact, a fraction of samples exhibit traces $\mI(t)$ that does not indicate a discernible trend towards equilibration, $\mI(t){\to} 0$ in the limit $t\to\infty$, at all. 
This point is also illustrated in Fig. \ref{f5}, second column. It shows an effective exponent
\begin{align}
    \beta(t)\coloneqq \frac{\partial \log\mI(t)}{\partial \log t}
    \label{e6}
\end{align}
that characterizes the time-evolution of $\mI(t)$ for an individual sample; the corresponding (logarithmic) distribution functions $\mP(\log \beta)$ are given in Fig. \ref{f6}. As is clearly illustrated by the data, the exponent distribution is logarithmically wide. 
Extremely strong sample-to-sample fluctuations have been observed also by other authors. As an interesting example, we mention \textcite{Doggen2018}: these authors use a machine-learning algorithm in order to distinguish localized from thermal samples. Within a regime of intermediate disorder values and system sizes the algorithm identifies the existence of two different sample types: those with insulating character and manifestly thermalizing others. 

\subsubsection{Restoration of self-averaging} 
The tails of $\mP$ are seen to be shrinking with increasing system size in Fig. \ref{f6}, if only exceedingly slowly. The general expectation is that at $W<W^*$ the imbalance, $\mI$, in interacting, disordered wires is self-averaging. The statement implies that in the limit of large systems, $L\gg 32$, the distribution $\mP$ acquires zero width, eventually. In other words, in the thermodynamic limit all samples, except for a small number of measure zero, are expected to thermalize with a relaxation behavior that is characterized by the same ($W$-dependent) exponents.

Due to computational limitations in system sizes and observation times, we can neither confirm nor dispute this expectation. If it is true, then all traces shown in Fig.~\ref{f8} will  undergo a  slow evolution so that they eventually collapse in the thermodynamic limit. 
However, also in this case it should not go unnoticed that for the range of system sizes traditionally studied in numerics, e.g. $L\lesssim 32$,  individual samples hardly ever show the typical behavior because the exponent distribution $\mP$ is so wide.

\subsection{Relation to previous work: creep and RG} 

For further discussion, we recall the RG-scenario outlined in the introduction. It assumes that samples fall into roughly two categories, thermalizing (ergodic) ``bubbles" and non-thermalizing ``grains". Fig.~\ref{f5}, lhs column supports this picture and allows rough quantitative estimates: the thermalizing fraction of samples comprises seven or eight out of ten samples at $W{=}1.5$, so  $\mfq(L{=}24,W{=}1.5){\approx}80$\%;  this fraction is rapidly decreasing for a larger disorder, e.g., we have only $\mfq(L{=}24, W{=}2.5){\approx} 50$\% at $W{=}2.5$ and $\mfq(L{=}24,W{=}5.0){\approx}0$. 

One can consider growing the sample size, e.g., by merging two samples of $L=24$, each. The RG assumes that the bubble and grain scenario continues to hold after merging and then predicts the evolution of $\mfq(L,W)$ with $L$ by postulating phenomenological growths rules~\cite{Vosk2015,Zhang2016,Dumitrescu2017, Dumitrescu2018, Goremykina2018, Morningstar2019}
Of interest to us is the general picture of the relaxation dynamics that take place immediately after merging two samples.  Specifically, at moderate to strong disorder, $W\gtrsim 2.5$, most samples behave like grains  since $\mfq<1/2$. Hence, even when doubling the sample size the most likely outcome is that a grain is added to a grain, so the relaxation dynamics after merging is extremely slow and does not support thermalization. Only upon growing the system size even further a  bubble will be added eventually, which can foster the delocalization process. Since a rather small bubble will need to equilibrate a large grain, thermalization is especially slow. An exponentially slow equilibration process that is enhanced by growing the system size is the hallmark of  ``creep"~\cite{Bera2017, Weiner19}.  In this sense, our numerical observations of large sample-to-sample fluctuations seen in Fig.~\ref{f5}, the slow flow seen in Fig.~\ref{f6}, and creep are all qualitatively consistent with the simplified RG-picture of thermalization through bubbles and grains. 
\footnote{ In recent numerical works, attempts have been made to substantiate the picture and identify correlated/entangled clusters in disordered samples~\cite{Loic2018,TomaszBubblePRB21, Hemery2022}. 
Since the data analysis in these papers underlies the assumption of a critical point near $W\approx3.8$, it remains to be seen to what extent the conclusions are affected once creep is taken into account.}

The alert reader will have realized that we adopted from the RG-concept its intuitive phenomenological building blocks, i.e. bubbles and grain, and left aside an important result of the reported RG-flow, which predicts a critical point separating a localized from an ergodic phase \cite{Goremykina2018, Dumitrescu2018, Morningstar2019}. 
We have neglected this part of the RG studies, because, within the parameter regimes investigated, we don't see support for a critical scenario provided by ``ab-initio" simulations of microscopic models. 
We mention two possible reasons for the discrepancy - other than reentrance behavior at a larger system size outside of our observation window. (i) The critical point might be situated at extremely large disorder values, as suggested by \textcite{Morningstar2022} and \textcite{SelsBathPRB22}. 
(ii) The phenomenological RG-equations are oversimplifications that miss relevant terms with a delocalizing effect.

\section{Conclusion}

Short quantum wires at relatively weak interactions and intermediate disorder strength can exhibit localization behavior: initial deviations from the equilibrium density are highly resilient against equilibration; a non-vanishing fraction of them may not, in fact, equilibrate at all. Understanding the fate of shorter samples with respect to their relaxation dynamics when successively growing their length is the central theme of many-body (de-) localization (MBL).
The numerical investigation we have presented in this work motivates three statements: 

(i) We follow how the relaxation dynamics of the sublattice imbalance $\aI$ and its fluctuations $\aF$ evolve over a large range of disorder values, reaching from below the clean bandwidth to four times its value. The temporal flow of the ensemble dynamics is conveniently parameterized by $\aSe$, which acts as a model for an internal (ensemble) clock. The usefulness of the ``internal clock" concept becomes most apparent by demonstrating that for a broad range of disorder values time traces $\aI(t)$ and $\aSe(t)$ can be collapsed to a single master curve - for fixed system size. The collapse works within the entire window of investigated disorder values, which implies the absence of a localization transition even when the disorder exceeds the clean bandwidth by a factor of four. 

(ii) We observe extremely strong sample-to-sample fluctuations: For systems with $L=24$ sites and moderate disorder non-ergodic samples (grains) coexist with highly ergodic samples (bubbles). Previous work has reported an extremely slow flow toward equilibration when growing the system size (``creep")~\cite{Weiner19}. In this work we have argued that the creep phenomenon is closely related to the strong sample-to-sample fluctuations here observed; the relationship has been established by borrowing ideas from a real-space renormalization group approach and the avalanche concept \cite{Thiery2017}.

(iii) While the existing evidence points to a thermodynamic limit that represents an equilibrating thermal phase, the flow towards this limit suggests the existence of subphases. Specifically, when the disorder strength reaches about two times the bandwidth, accelerated dynamics gives way to decelerated dynamics as indicated in Fig.~\ref{f1}. In this work, we present additional numerical evidence for this scenario, based on the observation that the evolution of $\mF(t)$ exhibits two regimes: At weak disorder, typical fluctuations are damped more strongly than average fluctuations; at larger disorder, it is the other way round. 

\paragraph*{Outlook.}
As an outlook, we express our belief that the neglect of creep in many, if not most, of the earlier computational studies of MBL, invalidates in parts the data interpretation that has been offered in these works, presumably often in crucial ways. At the time being it is too early writing an MBL review from the perspective of many-body de-localization (MBdL), i.e. ``creep''. Nonetheless, a firm ground has been laid that allows for a more careful analysis of the physical phenomena to be encountered. 
Many fascinating ideas have been expressed in the past, some of them formulated as mathematical theorems, some of them in terms of toy models, all of them making predictions that merit a careful computational test. Of course, creep will have to be included as a hallmark of delocalization physics in the data analysis - and no longer be ignored. 

The most important statement supporting the existence of MBL goes back to Imbrie~\cite{Imbrie2016,Imbrie-review2017}. 
While Imbrie's proof is believed to be rigorous, it also relies upon assumptions, e.g., concerning the spectral statistics of the sample  Hamiltonian. It will be interesting to see in future work, whether the lack of evidence for MBL in the XXZ-Heisenberg model is due to the disorder being still too weak, due to the proof being not fully complete yet, or due to a much simpler reason,  which is that the theorem does not apply to the XXZ-chain, because its assumptions are not met. 

Another open pressing question concerns the nature of MBdL in correlated disorder, such as represented, e.g., by the Andre-Aubry potential (AA). In our previous studies~\textcite{Weiner19} of charge-density relaxations, we did not detect a qualitative difference between correlated and fully uncorrelated randomness~\SB{(also see preliminary data in App.~\ref{appA3})}; similarly, also here we have no indication that the dominating physics is related to the effects of rare regions. Both observations together prompt the expectation that the AA-model exhibits creep, i.e., MBdL. If true,  the interpretation of cold-atom experiments in terms of the observation of MBL proper is challenged~\cite{Schreiber2015, Luschen2017, RispoliExp18}. 

One more crucial issue that merits closer scrutiny relates to the choice of the initial state used for time propagation, e.g., a Ne\'el state. In thermalizing phases, it is tempting to assume that the qualitative dynamics seen in a quench mostly reflect  properties of the phase rather than the initial state; \SB{confirming preliminary evidence is given in App.~\ref{appA1}}. To what extent this remains true also in marginally thermalizing situations requires further study.

We conclude with a caveat: In this article, we have adopted a manner of speaking that is established in the MBL community according to which a phase is ``equilibrating" or ``thermal" if the temporal evolution of the entanglement entropy is at large times faster than $\ln t$ and if simultaneously the sublattice imbalance and derived quantities in the thermodynamic limit decay to zero. 
It should be noted that the traditional meaning of equilibration implies that {\it all} local observables relax towards their equilibrium value, eventually. Whether the equilibrating/thermalizing phase(s) indicated in Fig.~\ref{f1} also satisfies such a stronger condition is a matter of ongoing research.

\section{ACKNOWLEDGMENTS}
We would like to thank J. Bardarson, I. Gornyi, A. Mirlin, M. Kiefer-Emmanouilidis, J. Sirker, and J. Zakrzewski for critical reading of the manuscript, and for their valuable comments, which improved our manuscript considerably. 
SB would like to thank S. Nandy for several discussions, and for an earlier collaboration on a closely related topic. FE expresses his gratitude to A. Rosch for an enjoyable set of conversations on the topic. SB acknowledges support from SERB-DST, India, through Matrics (No. MTR/2019/000566), and MPG for funding through the Max Planck Partner Group at IITB.
Funding from the Deutsche Forschungsgemeinschaft (DFG, German Research Foundation) through EV30/11-2, EV30/12-1, EV30/14-1, EV30/14-2 is gratefully acknowledged. IM acknowledges financial support from Prime Minister’s
Research Fellows (PMRF) scheme offered by the
Ministry of Education, Government of India.

\appendix

\SB{

\section{Further confirmation of the generality of the internal-clock concept}

In this section we provide additional evidence for the general applicability of the concept of the (ensemble) internal clock for synchronizing time traces obtained in different disorder regimes. 

\subsection{Different initial condition: Neel-block states
\label{appA1}}
\begin{figure}[b]
    \centering \includegraphics[width=1\columnwidth]{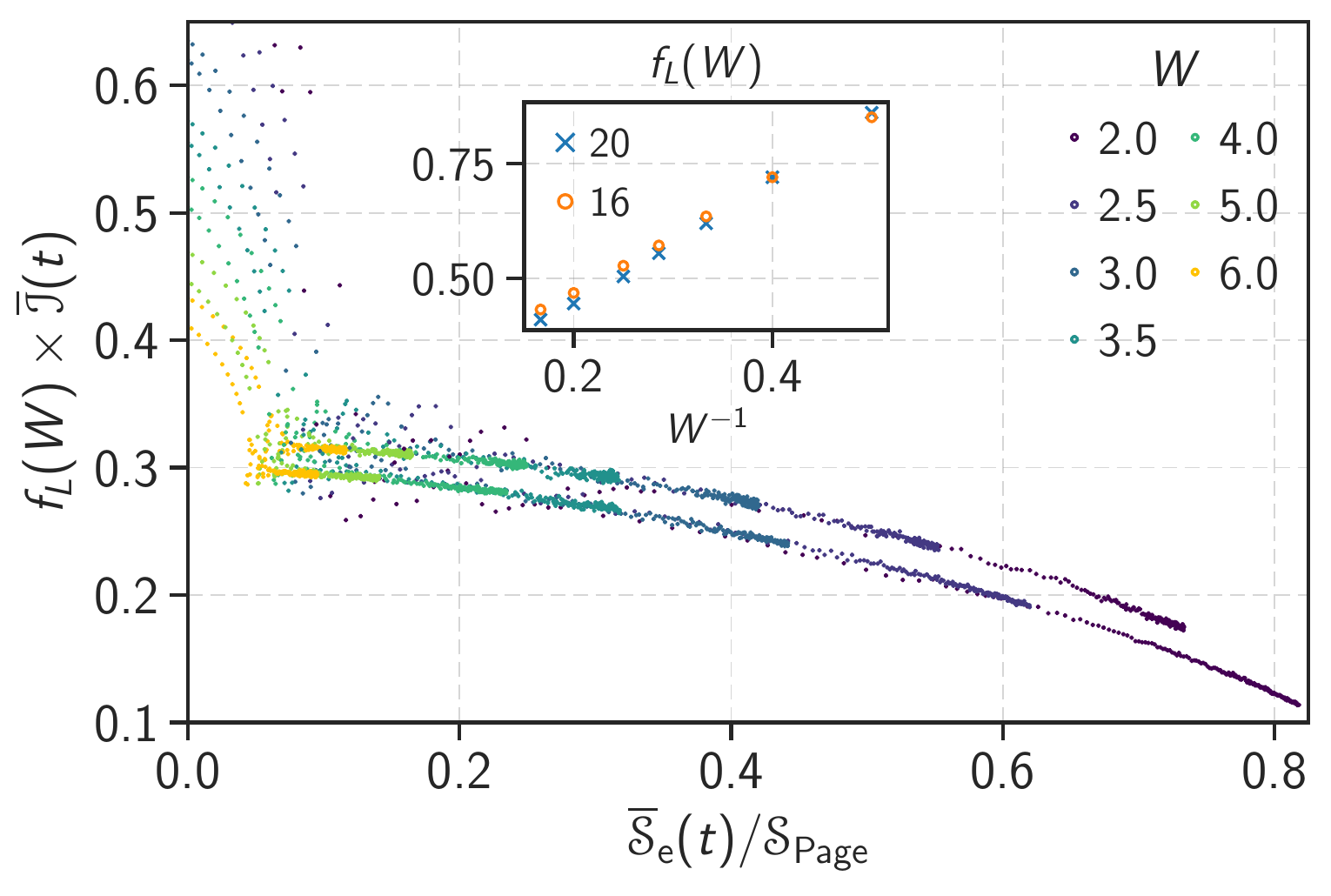}
    \caption{\SB{Time evolution of the imbalance, Eq. \ref{eA1}, after quenching a Ne\'el block state, $|\uparrow\uparrow\downarrow\downarrow\ldots\rangle$ for two system sizes $L=16,20$. The figure is fully analogous to Fig. \ref{f2}, which is for quenching a Ne\'el state. In both cases the data collapse is seen to be of a similar quality, underlining the generality of the internal clock as a means to synchronize time evolution in different ensembles. 
(Number of samples 200 ($W<4$) and 600 ($W>4$). }
    }
    \label{f13a}
\end{figure}
Synchronizing by using the evolving entanglement entropy as a measure of time also works for initial states different from the Ne\'el state. We demonstrate this by quenching from a Ne\'el-block state $|\uparrow\uparrow\downarrow\downarrow\ldots\rangle$, which similar to the Ne\'el state is situated in the middle of the many-body spectrum. For the block state, we define the imbalance in the following way:
\begin{equation}
    \aI(t) = \sum_{j=1}^L \ n_{4j-3}(t) + \ n_{4j-2}(t) - \ n_{4j-1}(t) - \ n_{4j}(t) \label{eA1} 
\end{equation}
 After quenching Ne\'el-block states, we observe an approximate collapse of all traces for different disorder values, see Fig. \ref{f13a}, similar to the case of the Ne\'el state, Fig. \ref{f2}. 
 

\subsection{Third observable: Correlation function \label{appA2}}
\begin{figure}[b]
    \centering \includegraphics[width=1\columnwidth]{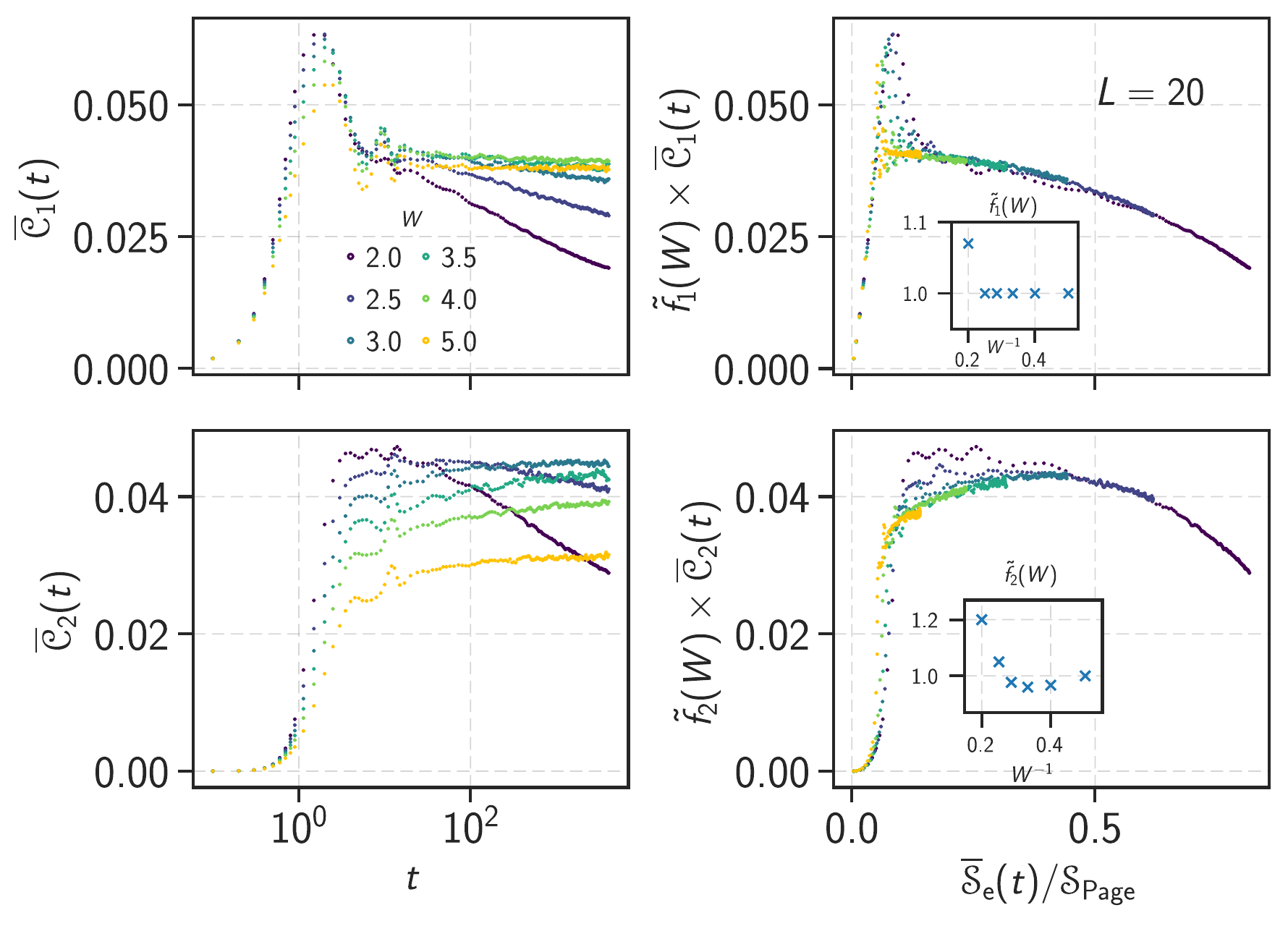}
    \caption{\SB{Time evolution of the nearest, $\overline{\mathscr{C}}_1(t)$ (upper row), and next nearest,  $\overline{\mathscr{C}}_1(t)$ (lower row), density correlations for system size $L=20$. 
    Similar to the imbalance, Fig. \ref{f2}, and the density fluctuations, Fig. \ref{fig:dephasing}, after employing $\aSe(t)$ as an internal clock and rescaling the ordinate by a scale factor $\tilde f_{1,2}(W)$ a (reasonably) good scaling collapse is achieved, see rhs column. 
   Fitting parameters: $\tilde{f}_1(W)=[1, 1, 1, 1, 1, 1.07]$ and  $\tilde{f}_2(W)=[1, 0.967, 0.96, 0.977, 1.05, 1.2]$ for $W$ ranging from 2.0 to 5.0. 
    (Number of samples: 200 at $W<4$ and 600 at $W>4$.) }
    }
    \label{f14}
\end{figure}
In order to demonstrate that $\aSe(t)$ potentially synchronizes a large set of observables, we here give a third synchronizing example, in addition to the average imbalance $\aI(t)$ and density fluctuations $\aF(t)$ considered in the main text. Specifically, we consider the density auto-correlation function defined as 
\begin{equation}
    {\mathscr{C}_l}(t) = \frac{1}{L-l}\sum_{i=1}^{L-l} C_{i, i+l}(t),    
\end{equation}
with $ C_{i, i+l}(t)=
    \langle \hat{n}_i \hat{n}_{i+l} \rangle - \langle \hat{n}_i\rangle \langle \hat{n}_{i+l} \rangle; 
$
$\langle \ldots \rangle$ represents expectation value in the state $|\psi(t)\rangle = e^{-iHt/\hbar} |\psi(0)\rangle$.
We consider ensemble averages, which restore the translational invariance: $\overline{\mathscr{C}}_l(t)$ is the ensemble averaged correlator where the overline stands for the ensemble average. 

Fig. \ref{f14} shows our results for nearest-neighbour and next nearest-neighbour correlations, i.e. $\overline{\mathscr{C}}_1(t)$ and $\overline{\mathscr{C}}_2(t)$. 
As one infers from the right-hand side column, a reasonable collapse is also achieved for this observable when using $\aSe(t)$ as an effective ensemble time. 

\subsection{Quasi-periodic potential\label{appA3}}
As a sanity check, we here demonstrate that synchronization with $\aSe(t)$ also works in situations of correlated randomness. 
To this end we adopt the Aubry-Andre model which is given by eq.(\ref{eq:H}) with $\epsilon_i=\frac{W}{2}\cos({2\pi\sigma i+ \phi})$, where, $\sigma = \frac{\sqrt{5}-1}{2} $ is the golden ratio and $\phi$ is a random number chosen uniformly from the range 0 to $2\pi$.
As seen in Fig. \ref{f15}, the quasi-periodic model shows a collapse of the imbalance trances similar to the situation with uncorrelated disorder model, Fig. \ref{f2}.
This strengthens the general applicability of the concept of a synchronizing interal clock. 

\begin{figure}[t]
    \centering \includegraphics[width=1\columnwidth]{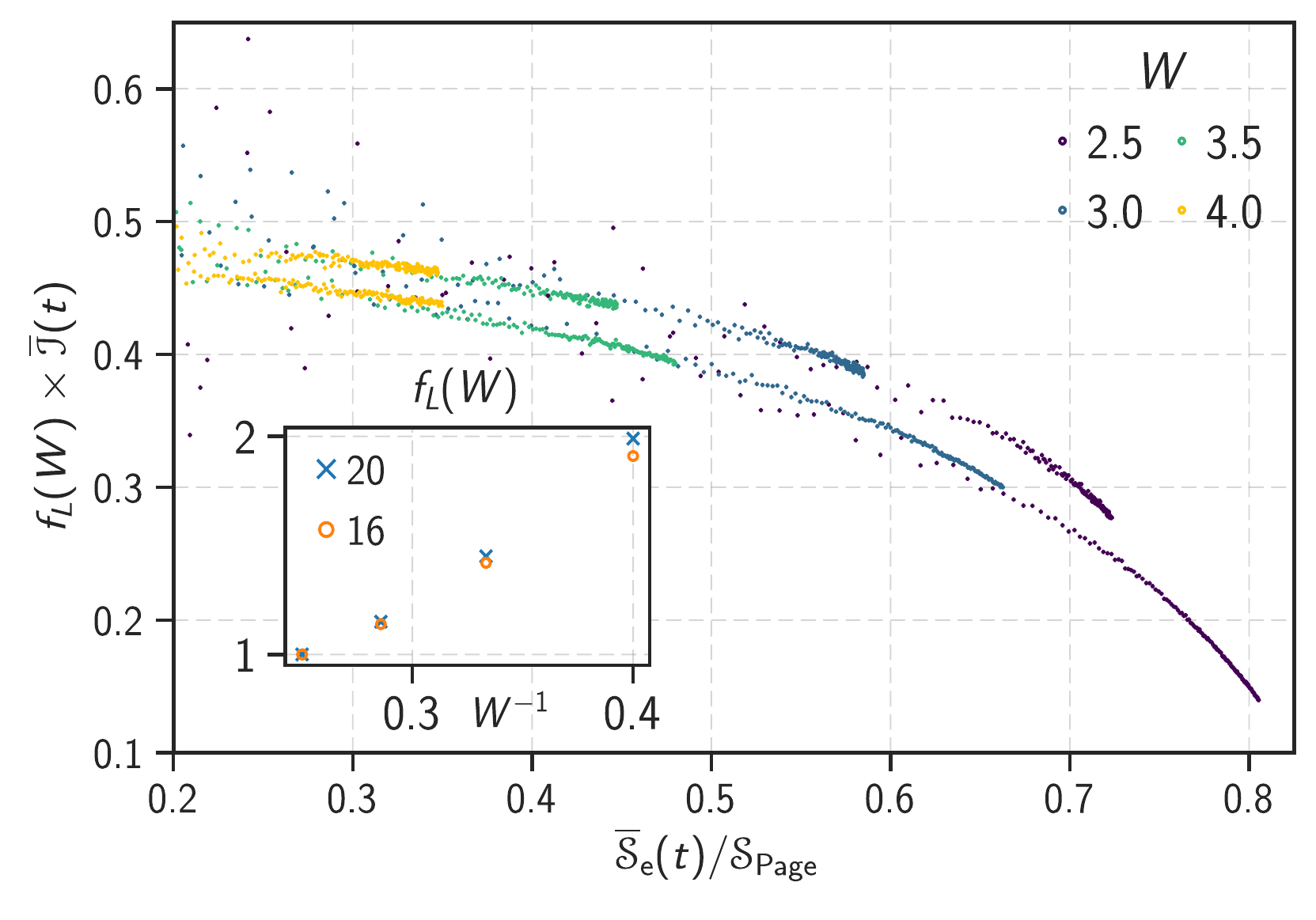}
    \caption{ \SB{Evolution of ensemble-averaged particle imbalance after quenching N\`eel state at system size $L=16,20$ for the case of correlated disorder (Aubry-Andre model). The figure is fully analogous to Fig. \ref{f2}, which features uncorrelated disorder. 
Inset: Scaling factor $f_L(W)$ as a function of $W$.
    (Number of disorder realizations: 600). }
}
    \label{f15}
\end{figure}
}

\section{Fitted parameters} 
The list of the parameters obtained from fitting power-law corrections, \eqref{e2} to the data given in Fig.~\ref{f4}. 
\begin{table}[!t]
\begin{tabular}{ l l | c c c c }
    \multicolumn{2}{c|}{$W$ \quad $L$} & $\aSe^\infty$ & $\aSe^\infty/\msr{S}_\mathrm{Page} $ & $c_0$ & $\gamma_L(W)$\\
    \midrule \midrule
        0.0	& 16 & 3.568(4)   & 0.707   & - & - \\
	& 20 & 4.504(21)  & 0.700 & - & - \\
	& 24 & 5.453(12)  & 0.697 & - & - \\
	& 26 & 5.929(6)   & 0.696 & - & - \\
\hline 
1.25 & 16 & 3.90(1) & 0.773 & - & - \\
    & 20 & 5.118(17) & 0.796 & - & - \\
    & 24 & 6.253(20) & 0.800 & - & - \\
    & 26 & 6.848(24) & 0.805 & - & - \\
\hline  
1.5 & 16 & 3.871(3) & 0.767 & - & - \\
    & 20 & 5.158(5) & 0.802 & - & - \\
    & 24 & 6.387(8) & 0.817 & 83(10) & 0.852(40) \\
    & 26 & 7.012(8) & 0.824 & 74(4) & 0.796(23) \\
\hline    
2.0 & 16 & 3.524(7) & 0.698 & 40(10) & 0.807(48) \\
    & 20 & 5.117(10) & 0.796 & 13.96(30) & 0.445(54) \\
    & 24 & 6.645(30) & 0.850 & 18.66(24) & 0.40(9) \\
    &26 & 7.476(9) & 0.878 & 19.68(13) & 0.369(3) \\
\hline 
2.5 & 16 & 2.928(1) & 0.580 & 19.32(7) & 0.653(1) \\
    & 20 & 4.450(5) & 0.692 & 10.95(10) & 0.362(4) \\
    & 24 & 6.307(5) & 0.807 & 13.13(4) & 0.269(1) \\
    & 26 & 7.372(10) & 0.866 & 14.58(8) & 0.239(2) \\
\hline    
3.0 & 16 & 2.371(4) & 0.470 & 10.75(80) & 0.545(8) \\
    & 20 & 3.497(11) & 0.544 & 9.61(41) & 0.356(7) \\
    & 24 & 5.187(10) & 0.663 & 9.96(19) & 0.226(7) \\
    & 26 & 6.70(20) & 0.788 & 10.87(17) & 0.166(10) \\
    \bottomrule
\end{tabular}
\caption{\label{t1} Parameters obtained from fitting the expression \eqref{e2} for $\aSe(t)$ to the data given in Fig. \ref{f4}a). Fits of the clean limit, Fig. \ref{f3a} right column, have also been included.
}
\end{table}

Details of the procedure are described in the main text; the ratio $\aSe^\infty/\mS_\mathrm{Page}$ is plotted in Fig. \ref{f4}(d). 
At weak disorder, the saturation values have been read directly from the original traces. At stronger disorder, $W>3.0$ the saturation behavior moves out of the time window where the ansatz \eqref{f5} would apply. 

To get a reasonable estimate of the error in the fitting parameters, we use the bootstrap method~\cite{YoungBS12, nrecipies}. 
This involves generating 100 new random sample traces around the original trace with a variance that is given by the sampling error. All these traces are then fitted with Eq.~\ref{e2} to get a spread with $95\%$  confidence interval in the fitting parameters $\aSe^\infty, \gamma_L(W)$.

\begin{figure}[b]
    \centering
    \includegraphics[width=0.95\columnwidth]{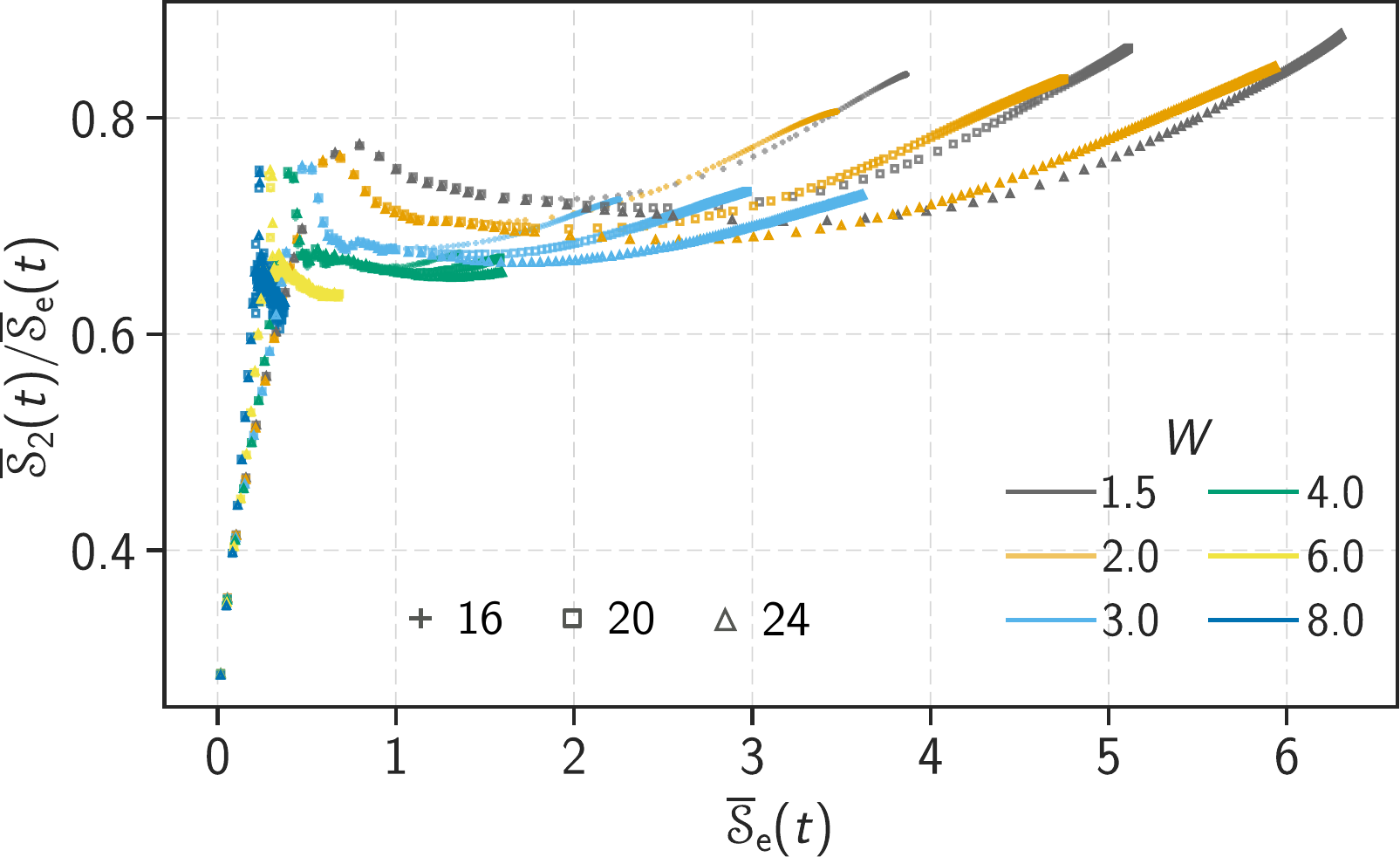}
    \caption{The time evolution of the (second) Renyi entropy in units of $\aSe$ in a broad range of disorder values, and system sizes $L=16, 20, 24$.}
    \label{f12}
\end{figure}
\section{$\mathscr{S}_2$ vs $\mathscr{S}_\mathrm{e}$}\label{app1}
To address the issue we depict in Fig. \ref{f12} the ratio of $\overline{\mS}_2/\aSe$. The ratio deviates from (roughly) a constant at short times and at long times, where system-size effects enter. We conclude that the entropy model for the internal clock has the best chance to work in an intermediate time regime where the ratio adopts a plateau-type behavior. Finite-size effects set in as soon as $\aSe(t)\propto \mS_\mathrm{Page}$ with a prefactor that shrinks with increasing $W$.

As a model for the internal clock, we have adopted the entanglement entropy $\mSe(t)$. Alternative choices are available, such as the generalized Renyi-entropies
\begin{align}
\mS^{(\alpha)} \coloneqq  (1-\alpha)^{-1}\ln \text{Tr}_\mathrm{A} \hat \rho_\mathrm{A}^\alpha 
\end{align}
with $\stwo=\mS^{(2)}$ and $\mSe=\mS^{(1)}$. Therefore, a question arises concerning the equivalence of different choices. 

\begin{figure}[b]
    \centering
    \includegraphics[width=1\columnwidth]{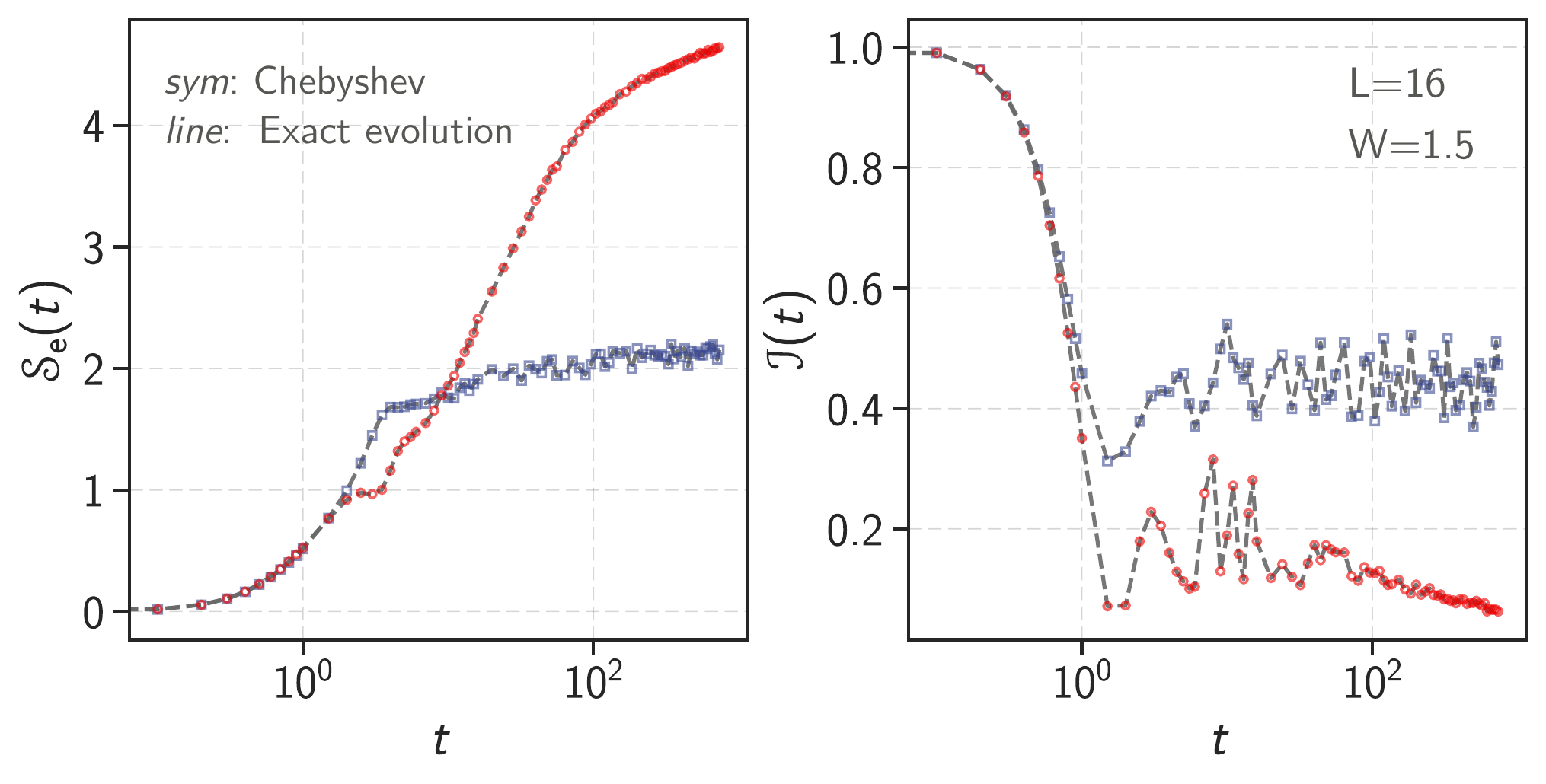}    
    \caption{Shows the time evolution of entanglement entropy and imbalance for both the exact and the Chebyshev method for two randomly chosen disorder realizations (red and blue) for $L=16$, and $W=1.5$.}
    \label{f13}
\end{figure}
\section{Convergence of time evolution}
\label{app:conv}
The validity of the conclusions presented in this work relies upon the reliability of our simulation data. 

To demonstrate the accuracy of the Chebyshev expansion as we have implemented it, we have performed a per-sample comparison with data from exact diagonalization (ED). Figure~\ref{f13} displays the result for two typical, i.e. randomly chosen 
samples. As seen there, within our simulation time $t\sim 10^3$ the traces are indistinguishable. The main reason for the usage of the Chebyshev expansion is that it allows treating systems larger than those affordable with ED, i.e. 
$L\gtrsim 16$ .

\bibliography{MBL}

\end{document}